\documentclass[preprint,12pt]{aastex}
\usepackage{verbatim}
\begin{document}
\makeatletter
\newenvironment{inlinetable}{%
\def\@captype{table}%
\noindent\begin{minipage}{0.999\linewidth}\begin{center}\footnotesize}
{\end{center}\end{minipage}\smallskip}

\newenvironment{inlinefigure}{%
\def\@captype{figure}%
\noindent\begin{minipage}{0.999\linewidth}\begin{center}}
{\end{center}\end{minipage}\smallskip}
\makeatother

\def\num1{${\cal D}_1$}
\def\mess{${{\theta_E}\over{\theta_1}}$}
\def\tetae{{$\tau_1 ({\rm days})$}}  
\def\br1br{${\cal R}_1^b ({\rm yr}^{-1})$}  
\def\r1d{${\cal R}_1^d ({\rm yr}^{-1})$}  
\def\rtb{{\bf ${\cal R}_{tot}^b ({\rm yr}^{-1})$}}  
\def\rtd{\bf{${\cal R}_{tot}^d ({\rm yr}^{-1})$}}   
\def\gc{globular cluster}
\def\kms{{\rm km/s}}
\def\rmeq#1{\eqno({\rm #1})}
\def\gl{gravitational lens} \def\gb{Galactic bulge}
\def\lc{light curve}
\def\ml{microlensing} \def\mo{monitor}
\def\pr{program} \def\mlmpr{\ml\ \mo\ \pr}
\def\ev{event}
\def\ex{expansion}
\def\fn{function} \def\ch{characteristic}
\def\bi{binary}  \def\bis{binaries}
\def\rd{Di\thinspace Stefano}
\def\bl{binary lens} \def\dn{distribution}
\def\pop{population}
\def\ct{coefficient}
\def\cc{caustic crossing}
\def\cs{caustic structure}
\def\mag{magnification}
\def\pl{point lens}
\def\dm{dark matter}
\def\rd{Di\thinspace Stefano}
\def\tots{track of the source}
\def\de{detection efficiency}
\def\det{detection}
\def\ob{observ}
\def\ol{observational}
\def\od{optical depth}
\def\ml{microlensing}
\def\mtm{monitoring team}
\def\mmtm{microlensing monitoring team}
\def\otm{observing team}
\def\mo{monitor}
\def\motm{microlensing observing team}
\def\los{line of sight}
\def\ev{event}
\def\by{binarity}
\def\ptb{perturb}
\def\sgf{significant}
\def\bis{binaries}
\def\sg{signature}
\def\bbl{binary-lens}
\def\kms{{\rm km/s}}
\def\rmeq#1{\eqno({\rm #1})}
\def\gl{gravitational lens} \def\gb{Galactic bulge}
\def\lc{light curve}
\def\ml{microlensing} \def\mo{monitor}
\def\pr{program} \def\mlmpr{\ml\ \mo\ \pr}
\def\ev{event}
\def\ex{expansion}
\def\fn{function} \def\ch{characteristic}
\def\bi{binary}  \def\bis{binaries}
\def\rd{Di\thinspace Stefano}
\def\bl{binary lens} \def\dn{distribution}
\def\pop{population}
\def\ct{coefficient}
\def\cc{caustic crossing}
\def\cs{caustic structure}
\def\mag{magnification}
\def\ppl{point-lens}
\def\bl{blending}
\def\mage{magnification}
\def\fsse{finite-source-size-effects}
\def\cc{caustic crossing}
\def\mp{multiple peak}
\def\lcf{lightcone fluctuation}
\def\wdf{white dwarf}
\def\pn{planetary nebula}
\vskip -.8 true in
\title{Orbital Motion During Gravitational Lensing Events}
\author{Rosanne Di\thinspace Stefano}
\affil{Harvard-Smithsonian Center for Astrophysics, 60
Garden Street, Cambridge, MA 02138}
\author{Ann Esin}
\affil{Harvey Mudd College, 301 Platt Blvd., Claremont, CA 91711}

\def\gl{gravitational lensing}
\def\Gl{Gravitational lensing}
\def\ml{microlensing} 
\def\Ml{Microlensing} 
\def\Et{Einstein angle} 
\def\et{$\theta_E$} 
\def\Er{Einstein radius}
\def\ev{event}
\def\vb{variable}
\def\vy{variability}
\def\sg{signature} 
\def\asec{arcsecond}

\begin{abstract}
Gravitational lensing events provide unique opportunities to discover 
and study planetary systems and binaries. 
Here we build on
previous work to explore the
role that orbital motion can play in both identifying and
learning more about multiple-mass systems that serve as gravitational lenses.  
We find that 
a significant fraction of planet-lens and binary-lens light curves 
are  influenced by orbital motion. 
Furthermore, the effects of
orbital motion extend the range of binaries for which
lens multiplicity can be discovered and studied.
Orbital motion will play an increasingly important role as
observations with sensitive  photometry, such as those made by the space missions
{\sl Kepler}, {\sl Transiting Exoplanet Survey Satellite, (TESS )}, and {\sl WFIRST} discover gravitational lensing events.
Similarly, the excellent astrometric measurements made possible by {\sl GAIA} 
will allow it to study the effects of orbital motion. Frequent observations, such as 
those made possible with the {\sl Korean Microlensing Telescope Network},
{\sl KMTNet} will also facilitate the study of orbital motion during 
gravitational lensing
events. Finally,
orbital motion will typically play a significant role in the characteristics 
of lensing events in which the passage of a specific nearby star in front of a 
background star can be predicted.  
\end{abstract} 

\section{Introduction}

\subsection{The Importance of Orbital Motion During Lensing Events} 

Many of the masses
 serving as lenses in gravitational lensing events are 
planetary systems, binaries, or higher-order multiples. As the lensing 
events proceed, the bodies constituting the lens system continue to orbit
each other. Orbital motion changes the magnification pattern as the
lens passes in front of the lensed source. When the event lasts longer
than the time required
for a significant change in orbital phase, the light curve may exhibit 
unique characteristics.

In this paper, we 
establish that a significant fraction of all light curves exhibiting
evidence of lens multiplicity should
also provide detectable signatures
of orbital motion. 
Our focus is on identifying the characteristics of 
those systems for which orbital motion
produces detectable features in gravitational lensing light curves,
The ``characteristics'' we consider include intrinsic properties
(mass and orbital properties), and also the placement of the
binary relative to us (distance and the orientation of
its orbit).  
For systems likely to exhibit the effects of orbital motion, 
we study the range of possible effects
and consider 
how these effects can be identified and studied.
Search for orbital motion will make  
lensing
observations more effective in discovering binaries and planetary
systems. These searches will become more successful as we move to an
era in which (1)~lensing events caused by nearby stars can be predicted
(L{\'e}pine
\& DiStefano 2012), (2)~the photometric and astrometric
precision of several space missions can be applied to lensing studies
({\sl Kepler} (Borucki
\& Koch 2012), {\sl TESS} (Ricker et al. 2014), 
{\sl WFIRST} (Yee et al. 2014), {\sl GAIA} (Proft et
al. 2011, Dominik \& Sahu 2000), and 
(3)~round-the-clock high-cadence sampling commences from the ground 
{\sl KMTNet} (Henderson et al. 2014) .

\subsection{Historical Background} 

Lensing light curves can display dramatic changes from the
point-lens form when the lens is a binary or planetary system. 
Early investigations of lens multiplicity focused on 
cases in which the event duration is so much shorter 
than the orbital period
that orbital motion produced subtle or negligible effects,
Indeed, evidence of rotation discovered so far
has been found in a handful of systems that have experienced modest
changes in orbital phase during the lensing event.  

\rd\ (2008) considered the case in which the orbital period is comparable to
or smaller than the event duration. In this cases, close approaches to a 
binary lens can produce light curves that are far more complex
than the most intricate light curves generated by a non-rotating
binary lens. The top panel of Figure~1 of that paper
shows the result of a head-on approach at low speed to a double brown-dwarf
binary. The event exhibits $6$ caustic crossings and $5$ additional
obvious peaks. Although the binary completed approximately two orbits
during the detectable portion of the event, the light-curve pattern
is not obviously periodic, but there is some evidence of quasiperiodic
behavior in  
 the low-magnification
wings of the light curve. 
It is not clear whether such a light curve  would have been identified as a 
candidate lensing event, particularly in the earlier incarnations of the 
lensing monitoring teams. Of course the approach between source and
lens in this case was special, in order that the source would pass over
regions with highly perturbed patterns of magnification. In the more
typical cases considered in this paper, the perturbations
are milder. Nevertheless, the complexity of the light curves and model fits
increases when the effects of orbital period, including possibly several
objects in orbit, orientation on the sky play significant roles.

\rd\ \& Night (2008) extended the study of orbital
rotation during events to the case of planets.
They showed earby planets orbiting in the habitable zones of their
stars, the ratio of orbital period to event duration
is also favorable for the detection of orbital motion.

To study the effects of a range of ratios between the event 
duration and orbital period, 
we utilize the concepts of the Einstein angle and the Einstein radius.  
When the source, lens, and observer are perfectly
aligned, the image of the source is a ring whose angular radius is
referred to as the Einstein angle, $\theta_E$. 
When the alignment is not exact, 
there are two images, whose separation is comparable to $\theta_E.$

When the angular separation between the distant source and intervening lens
is equal to $\theta_E,$ 
 the total magnification is $34\%$. 
The Einstein radius as projected onto the lens plane is just
$R_E=\theta_E D_L,$ where $D_L$ is the distance to the lens. 
\begin{equation}
R_E=0.6\, {\rm AU}\,  \sqrt{\Big({{M}\over{M_\odot}}\Big)
                        \Big({{D_L}\over{50\, {\rm pc}}}\Big)
                       \Big(1-\frac{D_L}{D_S}\Big)},
\end{equation}
where $M,$ is the total mass of the lens and $D_S$ is the distance to the
lensed source. 
The effects of lens binarity are most pronounced when $a,$ the
projected separation between the binary components is comparable to
$R_E.$ It is convenient to define $\alpha=a/R_E.$ 
Many of the binaries and planetary systems
observed as lenses so far tend to have values of 
$\alpha$ between roughly $0.2$~and~ $2.0$.

For a solar-mass lens located halfway between us and the Bulge,
$R_E \approx 5$~AU. 
\footnote{Most
lensing events have been discovered along directions to the Bulge. Thus,
because the  
value of $R_E$ is maximized at the point 
halfway
between the source and lens, the midpoint between the Earth and the
Galactic Center, with $D_L \approx 4$~kpc is often used for illustrative
purposes.}  
The orbital
period for binary lenses with $\alpha \approx 1$ can
therefore be several years.
In contrast, typical event durations have
been on the order of days, weeks, or months.  
Under these circumstances, the orbital phase should not change
very much during an event, and indeed, with hundreds of
binary-lens events identified, phase shifts have been measured
in only a handful of cases, and the values tend to be small
(Dominik 1998, Albrow et al. 2000, An et al. 2000). 

Dominik (1998) considered the effects of orbital rotation, including
cases in 
which either the lens or source is a binary. He found that, of these effects,
binary-lens rotation produces the most significant
light-curve perturbations. He identified an acceptable model 
fit to the DUO\#2 event
that includes the effect of rotation, but the phase change during the event was
small, since the Einstein radius crossing time was $6.4$~days, while the binary
orbital period was $85.4$~days.   
Ioka et al. (1999) found that Kepler rotation should
be significant for binary lenses in the Magellanic Clouds, although orbital
effects were not needed to fit MACHO LMC-9, a binary-lens event.
Dubath et al. (2008) considered fast-rotating lenses, but 
miscomputed the light curves by not
including the important non-linear effects associated
with binary lenses.  

More recently, 
Penny et al. (2011a) carried out a simulation of microlensing 
events produced by the Galactic binary population, and also
Penny et al. 2011b computed the detection efficiency for
orbital effects. 
Unsurprisingly, they find that 
averaged over all relevant parameters (distances, binary periods and 
separations, relative velocities, etc.), the detectable fraction is 
considerably lower than 0.1\%\footnote{In fact, their values should be 
considered to be 
upper limits, since their criterion for orbital-motion detection was simply 
a failure to obtain a good non-rotating binary lens lightcurve fit.}   This 
is of course consistent with the 
general consensus that such events are very rare, which is why the effects
of orbital motion in microlensing were largely ignored for so long.  More
importantly, Penny et al. explored the effects of different microlensing 
parameters on the rate of orbital motion detection.   
However, because their study focused on population 
averages, the interpretation of their results is not always straight-forward,
since it depends on the authors' choices for the distributions of critical
parameters (e.g. lens distances, periods, etc.). 
In any case, binary population calculations are complicated by the 
need to incorporate 
stellar and binary evolution as well as planetary systems. Calculations
conducted to calculate the rate of Type Ia supernovae,
for example, give different results when performed by different groups
(Nelemans et al. 2013).
Furthermore, there is evidence that
even the ``initial conditions'', corresponding to the range of
orbital periods, masses, mass ratios, and eccentricities in primordial
binaries, need to be better established 
(Moe \& Di\thinspace Stefano 2013a, 2013b).

In this paper we therefore take a different approach. We identify the
sets of system and event parameters that allow orbital motion to 
significantly influence the form of the light curve.
We study the effects. We then consider
a range of physical and mathematical symmetries that identify 
a huge range of systems expected to produce similar light curves.
Some of these systems have low mass stars, others are high-mass
binaries. Some have short orbital periods, others orbital periods of 
hundreds of days. Some are near to us, and some ar close to the source
star. 
We thereby show that the effects of orbital motion are
likely to be ubiquitous. We show that a calculational method, developed by
Guo et al. (2011), is
needed to search for and consistently find evidence for it, and to reliably
measure 
 binary 
orbital periods.

\subsection{Plan of the Paper} 

In \S 2 we identify the regions of the parameter space for 
planetary and
binary systems in which orbital motion should influence the lensing
signatures. 
We find that for some sets of event parameters,
the effects of orbital motion can be detected even for
binary lenses with orbital periods of tens to hundreds of days.
Furthermore, the effects of lensing and the binary nature of the lens
can be more detectable when the light curve carries information about
orbital motion. We illustrate these results by considering a binary 
located at a fixed distance $D_L$ from the observer,
computing 
the effects of rotation for a range of orbital
separations. 
 
In \S 3 we turn to a set of five instructive examples with a broad range
of 
physical properties to demonstrate 
the connection between the rotating geometry of the lens and the
lensing signature.  
Although the number of systems we consider is relatively small, we show that 
each represents a large set of binaries.
Section~4 is devoted to our conclusions. We find
that a large fraction of those planetary systems and binaries for
which the lensing event provides evidence of a companion, the light
curve is also significantly influenced by orbital motion. Furthermore,
the effects of orbital motion extend the range of values of the orbital 
separation 
for which lens multiplicity can be discovered, 
and teach us more about the multiple system
that served as a lens. 
We conclude with a discussion of 
how observational studies can
identify more events exhibiting orbital motion.

\section{The Effects of Orbital Motion}

\subsection{Detecting Evidence of Binarity: Static Lenses}

We start by considering a point-like lens of mass $M.$
The lens position defines the center of a set of circular
isomagnification contours projected onto the sky. 
If we define $u$ to be the projected distance between the lens and the 
background source, expressed in units of the Einstein radius, $R_E,$ the
magnification depends only on the value of $u$: 
$A[u(t)]=(u^2 + 2)/(u \sqrt{u^2 + 4})$.  
Because the
lens has no structure, all angles of approach are equivalent, and  
the lensing light curve is symmetric about a
peak magnification that occurs at the time of closest approach between 
the lens and a background source of light.
Lensing model fits to the light curve provide the value of the distance of closest approach,
the time of closest approach, the baseline magnitude, and a
time duration $\tau_E,$ defined to be the time required for the lens-source distance to
change by $2\, R_E.$ The value of $\tau_E$ is the only measurable quantity
that depends on the lens mass, $M.$ Measuring $\tau_E$
does not, however,
 provide a unique measure of the lens mass, because
 value of $\tau_E$ is also
determined by  
the distances to the lens and source ($D_L$ and $D_S$, respectively), 
and the value $v$ of the relative transverse speed. The lensing
solution is therefore highly degenerate.

Lens binarity breaks the axisymmetry, and expresses itself 
through deviations in the isomagnification
contours. 
When the 
path of a distant star passes behind regions that deviate from the
point-lens pattern, the lensing light curve exhibits distinctive
features. Because these features introduce a physical
scale, they can help to break the degeneracy discussed above. For example,
sometimes finite-source-size effects can be detected when
binary-lens effects introduce short-lived deviations; these
can allow us to measure
the size of $\theta_E,$ and to thereby 
directly relate the mass of the lens to its
distance from us.

The
spatial structure of the deformations, and therefore the
shapes of the light curves depend on the binary's physical 
characteristics. For static binaries the two key quantities
are the mass ratio $q$ between the binary components and the 
projected orbital separation, $a$. We define $q=M_2/M_1,$ where $M_2<M_1$.
Lensing is not directly sensitive to the 3-dimensional orbital
separation, but instead depends on the value of the $a,$ the
orbital separation projected onto the plane of the sky. 
The effects are determined by the ratio $\alpha= a/R_E.$  
\begin{figure}[] 
\epsscale{.75}
\plotone{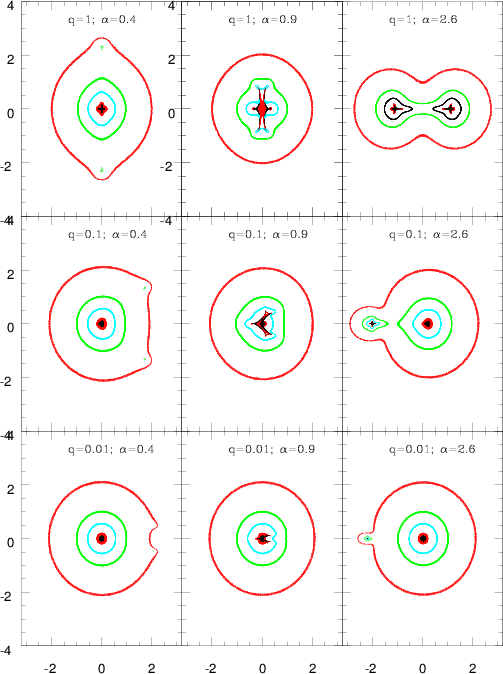}
\vspace{-.2 true in}
\caption{\label{binary}{\bf Isomagnification regions for 9 binary lenses.}  
Outer contour (red): A=2.05. Proceeding inward, $A=1.34$ along the green
contours, and $A=2$ along the cyan
contours. 
Inner points in red (black) mark regions in which the 
magnification is larger than 5 (10). 
The value of $\alpha$ increases from left to right and the value of $q$
decreases from top to bottom. \rd\ (2012) considers smaller values
of $\alpha.$   
}  
\end{figure} 


For a given value of $\alpha,$ the effects of binarity on the isomagnification
contours are small for values of $q$ in the range typical of planets,
but the linear dimensions of the regions in which there are 
significant perturbations increase with increasing $q$. 
Some examples are shown in
Figure \ref{binary}, with the patterns in the top, middle, and bottom rows
 corresponding to
$q=1,, 0.1,$ and $0.01,$ respectively.

For a given value of 
$q,$ the size of the region in which there are significant 
perturbations is typically largest for $\alpha \sim 1.$
 As $\alpha$ 
increases above unity, 
the isomagnification contours begin to mimic the contours expected
from two separate lenses. As $\alpha$ decreases below unity, the
region with large photometric deviations from the point-lens form
shrinks. 
The progression from small to large $\alpha$ can seen by scanning from left
to right in each row of Figure 1.

Because the deviations shrink for smaller values of $\alpha,$ most
previous investigations have not explicitly considered small $\alpha.$  
Nevertheless,
 at small $\alpha,$ there
is a region of perturbations at a distance roughly equal to $1/\alpha$ from
the center of mass (Di\thinspace Stefano 2012, Di\thinspace Stefano \& Night 2008). 
This is illustrated in the left-hand panels of Figure~1 and is 
addressed in detail for planets in \rd\ 2012. We also include examples of 
small-$\alpha$ binaries in this paper.

\subsection{Relating the orbital period to the lensing parameters}

The value of the orbital period, $P,$  
depends on the value of the total system mass, $M,$ which is also the
mass that enters into the expressions for the 
Einstein radius, $R_E$, and for the Einstein diameter
crossing time, $\tau_E$. 
The value of $P$ also depends on the size, $a_s$ of the semimajor axis.
If we define $\alpha_s = a_s/R_E,$ then
\begin{equation} 
P = 1~{\rm yr}~\, \Big(\alpha_s\Big)^{\frac{3}{2}}\, 
                  \Bigg({\frac{R_E}{AU}}\Bigg)^{\frac{3}{2}}\,
                  \Bigg(\frac{M_\odot}{M}\Bigg)^{\frac{1}{2}}  
\end{equation} 
It is not, however, $\alpha_s$ that determines the shapes of the 
isomagnification contours. Instead, it is the instantaneous value of 
the projected separation, expressed in units of $R_E$: $\alpha=a/R_E$.
In general $\alpha=\alpha(t)=f(t)\, \alpha_s.$ For eccentric orbits
viewed nearly face-on, $f(t) > 1$ near apastron. Near periastron, or for
inclined orbits $f(t) < 1$.
It is convenient to express $P$ as follows.

\begin{equation}
\label{period}
P = 170\, {\rm days}\, \alpha_s^{3/2} \Bigg[
{{M}\over{M_\odot}}\, \Bigg(\frac{D_L}{50\, {\rm pc}}\Bigg)^3 (1-\frac{D_L}{D_S})^3
\Bigg]^{1\over 4}
\end{equation}

In order for the effects of orbital motion to play a role in 
altering the shape of the light curve, the binary axis must rotate
significantly while the event is in progress.
We must therefore compare the time scale of the event to the 
orbital period.  
The time scale for photometric events is set by the value of 
$\tau_{E} = 2\, R_{E}/v,$ where $v$ is the relative transverse velocity
of the source in the frame of the lens. 
\begin{equation}
\label{tauE}
\tau_{E} = 41.6\ {\rm days} \Bigg({{M}\over{M_\odot}}\Bigg)^{{1}\over{2}}
             \Bigg({{D_L}\over{50\, {\rm pc}}}\, (1-x)\Bigg)^{{1}\over{2}}
             \Bigg({{50\, {\rm km/s}}\over{v}}\Bigg),
\end{equation} 
which combined with Eq. (\ref{period}) gives us the timescale ratio
\begin{equation} 
\label{ratio}
{{\tau_{E}}\over{P}}= {{2.73}}\, 
\Big({{0.2}\over{\alpha_s}}\Big)^{{3}\over{2}}\, 
    \Big({{50\, \rm{km/s}}\over{v_T}}\Big)\,  \Bigg[
{{M}\over{M_\odot}}\, {{50\, {\rm pc}}\over{{D_L}\, (1-x)}}
\Bigg]^{1\over 4}.   
\end{equation}
\begin{figure}
\epsscale{.5}
\plotone{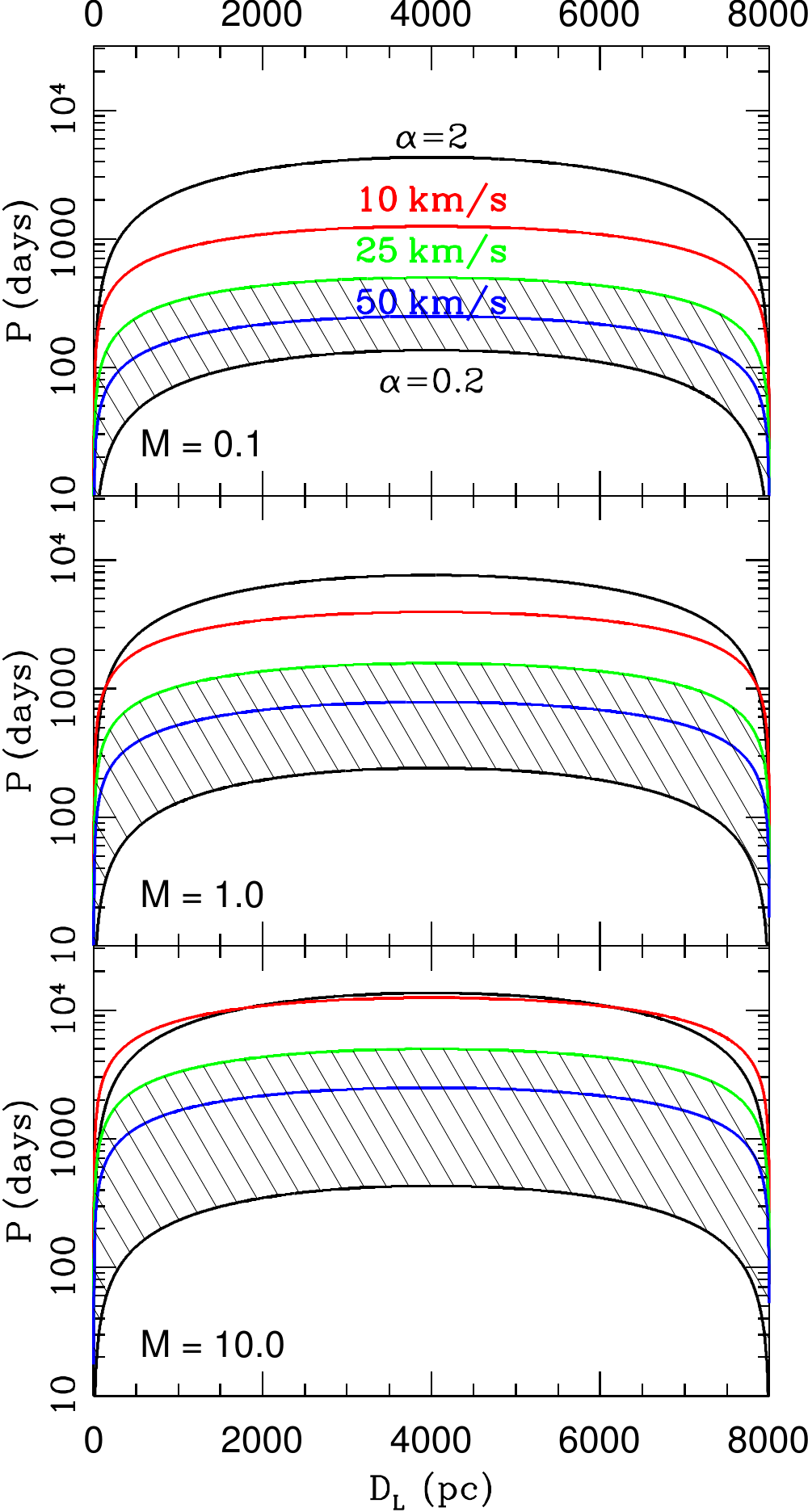}
\vspace{-.1 true in}
\caption{\label{prange}
Ranges of orbital periods, as a function of lens distance, for which
lens binarity can be reliably detected;
we take $D_S=8$~kpc.
{\sl Top panel:} $M=0.1\, M_\odot$;
{\sl Middle panel:} $M=1.0\, M_\odot$;
{\sl Bottom panel:} $M=10.\, M_\odot$.
In each panel, $P-D_L$ relationship shown in the uppermost (lowermost) black curves
is computed by using Eq.~3 for $\alpha=2$ ($\alpha=0.2$), the approximate range of
$\alpha$ for which lens binarity is detectable, even without orbital motion.
The colored curves show the $P_{max}-D_L$ relationship (Eq.~6), each for
a specific
value of the relative transverse velocity
({\sl red (top)}: $10$~km~s$^{-1}$;
 {\sl green (middle)}: $25$~km~s$^{-1}$;
 {\sl blue (bottom)}: $50$~km~s$^{-1}$).
The region below each of these curves corresponds to the region within which
the effects of orbital rotation significantly changes the light curve properties
(Eq. [\ref{Pmax}]).
For illustrative purposes,
the hatched region shows (for the
specific case of $v = 25\,{\rm km/s}$)
the intersection of the region
where binarity is potentially detectable (even without orbital motion)
with the region where orbital motion plays an important role and can
significantly deform the shape of the binary-lens light curves.
}
\end{figure}

\begin{figure}[h]
\epsscale{0.5}
\plotone{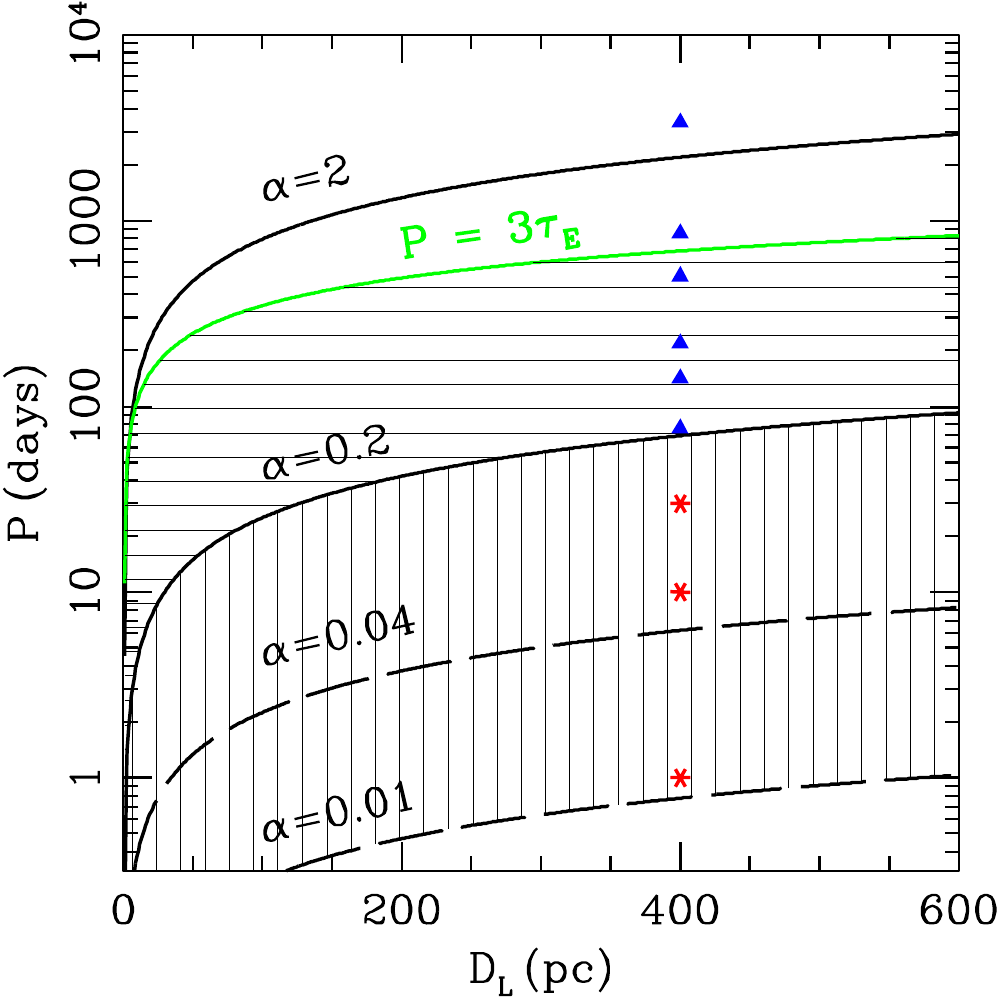}
\vspace{.75 true in}
\caption{\label{prange.small} Maximum and minimum binary periods (as in Figure
\ref{prange}) for nearby lenses with total mass $1M_{\odot}$ and relative 
velocity $v=25\,{\rm km/s}$.  The region with horizontal shading is the 
same as the shaded region in Figure \ref{prange}; here binary rotation causes
very significant changes to the lightcurve.  The binary lenses with periods 
in the region with vertical shading will show detectable rotation signatures
in microlensing lightcurves observed with sensitivity better than the 
present $\sim 1\%$.  The lightcurves for the lensing events produced by
systems labeled as blue triangles and red stars are shown in Figures 
\ref{lcperiods} and \ref{lcresiduals}.}
\end{figure}

\subsection{Detecting Evidence of Orbital Motion}

The timescale ratio, ${{\tau_{E}}/{P}}$, 
can indeed be very small (i.e. rotation is unimportant) for
large values of $\alpha$, $v$, and $D_L(1-x)$.
However, in this section we demonstrate that the value of 
$\tau_E/P$ large enough to produce detectable effects
for a wide range of binary-lens parameters.
A useful criterion is that the binary should execute an 
orbit during an interval of time equal to $3\, \tau_E$. This roughly
corresponds to completing an orbit during the  interval when the 
magnification is greater than $2\%.$ Today's ground-based missions
can, in principle, track the light curve through even lower values of
the magnification\footnote{Blending of light from the lensed source with light 
from the lens or from other background sources can make the detection of
low-magnification effects more difficult.}. Space missions can do even 
better.     
In data taken by the {\sl Kepler} mission, 
{\sl TESS}, or {\sl WFIRST},   
an event is potentially detectable during the interval when the lens-source
separation is as large as $10\, R_E;$ larger orbital periods would still
produce significant deviations from the point-lens light curve shape.

Imposing the condition $P \lesssim 3 \tau_E$ and using Eq. (\ref{period}) 
to eliminate $\alpha_s$ in favor of $P$, implies a maximum orbital 
period, $P_{max}.$ 
\begin{equation}
\label{Pmax}
P_{max} = 125\, {\rm days}\, \left(\frac{50\,{\rm km/s}}{v}\right) 
\left[{{M}\over{M_\odot}}\,\frac{D_L\,(1-x)}{50\, {\rm pc}}\right]^{1/2}. 
\end{equation} 
Note that orbital effects can be detected even when the orbital period
is somewhat longer. 

Figure 2 illustrates the range of orbital periods for which binary
and rotational effects are potentially detectable, as a function of distance $D_L$
to the lens. Each panel corresponds to a specific value of the lens mass,
ranging from $M=0.1\, M_\odot$ in the top panel to $M=10\, M_\odot$ in the
bottom panel. In each panel, there is an upper and lower black curve
showing $P$ versus $D_L$ for, respectively, $\alpha=2$, $\alpha=0.2$. This range of
$\alpha$ roughly corresponds to the range over which lens binarity
can significantly affect the shape of the light curve, even for static lenses. 
Thus, if a lens lies between
these two curves, lens binarity is potentially detectable, 
even without the effects of
orbital motion.
It is important to note that, in the static-lens case, not every detected lensing
light curve exhibits detectable evidence of binarity, even for systems 
for binaries in the 
region between $\alpha=0.2$ and $\alpha=2.$    
Instead, depending on the sensitivity and cadence of the observations,
there is a finite probability, often small, that lens binarity will be detectable.
The probability of detecting binarity is 
proportional to the probability that the source track will
pass behind a region with isomagnification perturbations, 
and is therefore proportional to the linear size 
of the perturbed regions.  
The probability increases 
with larger values of the mass ratio $q$,
and for values of $\alpha$ close to unity. 

 The colored curves in each panel plot $P_{max}$ versus $D_L$.
Each curve corresponds to a particular value of the relative velocity, indicated
by the curve's color (red: $10$~km~s$^{-1}$; green: $25$~km~s$^{-1}$; blue:
$50$~km~s$^{-1}$). For all systems located {\sl beneath} a curve of a given color
(whether or not $\alpha>0.2$), 
rotational motion is potentially detectable, since $\tau_E/P>3.$ 

Note that in each panel, the region between the green curve ($P_{max}$ for 
$v=25$~km~s$^{-1}$) and the bottom black curve ($P$ for $\alpha=0.2$) is
cross hatched. For binaries in this cross-hatched region, there is a significant
probability that the light-curve perturbations associated with binarity
will exhibit rotational effects. 

Rotational effects can lead to the repetition of deviations, making
it possible to reliably identify small deviations whose significance
would otherwise be unclear. Thus, the region in which binarity can
be detected may be extended to smaller values of $\alpha$ by the effects
of orbital motion. This is illustrated in Figure 4.  

Figure \ref{prange.small} is a blow-up of the small $D_L$ region of Figure 
\ref{prange}, for a lens mass of $1\, M_\odot.$ 
The region with horizontal shading above the curve for $\alpha=0.2$ is the 
same as the shaded region in Figure \ref{prange}.   The region with vertical 
shading shows the effects of extending the systems of interest to those 
with lower values of $\alpha$.   For these small values of $\alpha$, 
the expected photometric deviations from a single lens lightcurve form
are tiny.  However, if they can be detected (with space observations or 
improved ground-based photometry), the orbital motion 
can be so rapid that the interval during which the event occurs 
encompasses many orbital periods producing strong rotation signature (see 
Figure \ref{lcresiduals}).

\subsection{Effects of Orbital Motion}
\begin{figure}
\epsscale{1.0}
\plotone{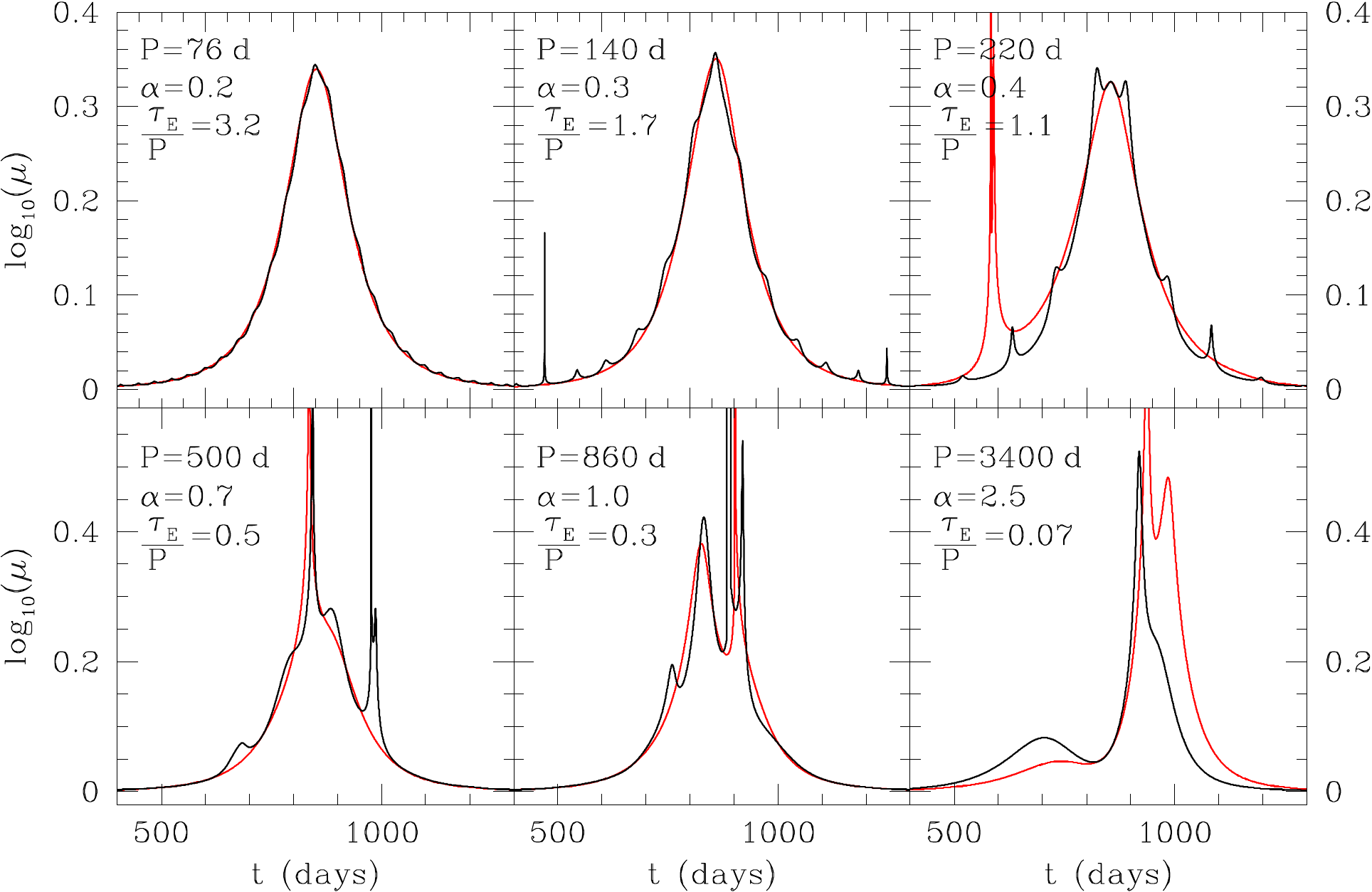}
\vspace{-.2 true in}
\caption{\label{lcperiods} Shown in black is the 
magnification as a function of time for 
microlensing events produced by binary lenses with periods
corresponding to the blue triangles in Figure \ref{prange.small}.  The
total lens mass, mass ratio,
distance, relative velocity and impact parameter are $1
M_{\odot}$, $0.8,$ 400\,pc, 25\,km/s and $0.5 R_E$ in all cases.  The
qualitative changes caused by the binary rotation are obvious for the
four lowest period values. Note that periodic modulation seen for
$P=76\,{\rm days}$ is at 2\% level, i.e.  detectable with present
microlensing searches.  The lightcurves for the longer period binaries
still differ significantly from one would see in the limit of no rotation.
This point is illustrated by the red curves, which are calculated for 
the same (but non-rotating) system and impact parameter; the trajectory 
phases were matched at the moment of closest approach.} 
\end{figure}

\noindent {\bf Light Curve Morphology:}
The effects of orbital motion can be predicted by studying the magnification
patterns shown in Figure \ref{binary}. Events are produced 
when a source 
track passes
behind the region near the Einstein ring of the lens. When the lens is
rotating, the magnification pattern rotates and/or oscillates. 

First consider a case in which the binary is face on.
This is an important case, since lensing is sensitive to binarity in
face-on orbits, while transit and radial-velocity methods are not.
Thus, lensing studies can provide important information about binaries
and planetary systems that complements what we learn through other methods.  
For a face-on orbit, the effects or orbital motion can be the following.
A source track which, for a static lens, would have passed behind no
distinctive features, will now  have the features rotate into its path.
In general, the number  of distinctive features, such as caustic crossings and
peaks and dips in the magnification,  is increased
by binary rotation. 

Edge-on orbits are also accessible to lensing studies. In this case, the 
pattern of isomagnification contours doesn't rotate, but instead it changes
because the projected separation between the binary components 
changes with orbital phase.  As the value of $\alpha$ increases and
decreases, the
magnification pattern oscillates. We can see this in the rows of
Figure 1, by considering the sequence from left to right and then from
right to left. 

In general, an orbit will have an arbitrary orientation on the sky, and
both rotational effects and the effects of changing values of $\alpha$ will
influence the light curve. Eccentric orbits, whatever their orientation, also
pass through a range of values of $\alpha$ as the orbit proceeds.   
Thus, whatever the orbital orientation or eccentricity,  
a common feature of orbiting-binary light curves is that they exhibit more structure
than is typical for static binary-lens light curves. It is important to note that,
even if the orbital period is too long for the light curve deviations to repeat, 
light curve model fits can nevertheless be used to derive the value
of $P.$

\noindent {\bf Repetitions:}
For events that are detectable 
over several orbital periods, we should be able to determine that the
signal is periodic and to measure the period. 
Tests for periodicity are complicated by (1)~the time dependence of the
overall magnification, (2)~the shape of typical isomagnification
contours, and (3)~the combination of translational and orbital motion.
To test for periodicity, we
have conducted a Lomb-Scargle analysis on each light curve.
To minimize the effects of trends in the overall magnification, 
we identify the best fitting point-lens light curve, and subtract
this from the observed light curve. To minimize the effects of the 
linear structures displayed in many binary-lens isomagnification
contours (which can produce two significant perturbations 
per orbit), we multiply measured periods by a factor of $2$.
(This is not necessary for small values of $q$.)     
It is more challenging to correct
for the contribution of translational motion.
In this paper we simply show the limitations of the standard approach, and
develop an improved method in a separate paper (Guo et al.\ 2011). 

\medskip


To illustrate the effects orbital motion has on microlensing lightcurves we
have chosen nine representative binary lensing systems shown as blue triangles 
and red stars in Figure \ref{prange.small}.  
Each binary has the same total mass ($1~M_\odot$) and the same mass ratio
($q=0.8$). 
Thus, they could be double M-dwarf systems, double white-dwarf
systems (double degenerates), or  
white-dwarf/M-dwarf binaries.   
Each binary is located $400$ pc from Earth.   
These values of $M$ and $D_L$ produce $R_E = 1.7$\,AU and $\theta_E = 
R_E/D_L = 4.2$~mas. To generate light curves, we set
$v=25$~km/s, producing a value for the Einstein diameter crossing time of 
$\tau_E= 234$~days, for all the systems and kept the distance of closest
approach at $0.5\, R_E.$  In every case, we assumed face-on circular orbits.  
The only difference among our nine systems is the orbital period (or orbital 
separation).  

Figure \ref{lcperiods} shows (in black) the light curve for each system 
represented by a blue dot in Figure \ref{prange.small}, in order of 
increasing period. A random orbital phase was chosen at the start of each 
computation. Also shown in red in each panel is the binary-lens light curve 
that would have
been obtained if there were no orbital motion. To make a meaningful
comparison, we matched the trajectory phases at the moment of
closest approach

We describe each panel, starting with the smallest value of $\alpha$ and
working toward higher values. On the top left is the panel for 
$\alpha=0.2,$ yielding an orbital period of $76$~days.   
The value of $\tau_E$ is $3.2$ times the orbital period. Thus, 
roughly $10$ orbital periods will pass during the time when the 
magnification is greater than $2\%$. By comparing the 
black (with rotation) 
and 
red (no rotation)
curves we can clearly see the periodic modulations. 
However, the magnitude of the signal is low, because the distortion 
of the isomagnification
contours for $\alpha=0.2$  
is slight.  
Nevertheless,   
applying the Lomb-Scargle analysis as described above, we identify 
a period at a high level of significance. The value we derive,
after correcting by the factor of $2$, is $70$ days, differing by less than 
10\% from the true period of $76$ days.
\begin{figure}
\epsscale{0.5}
\plotone{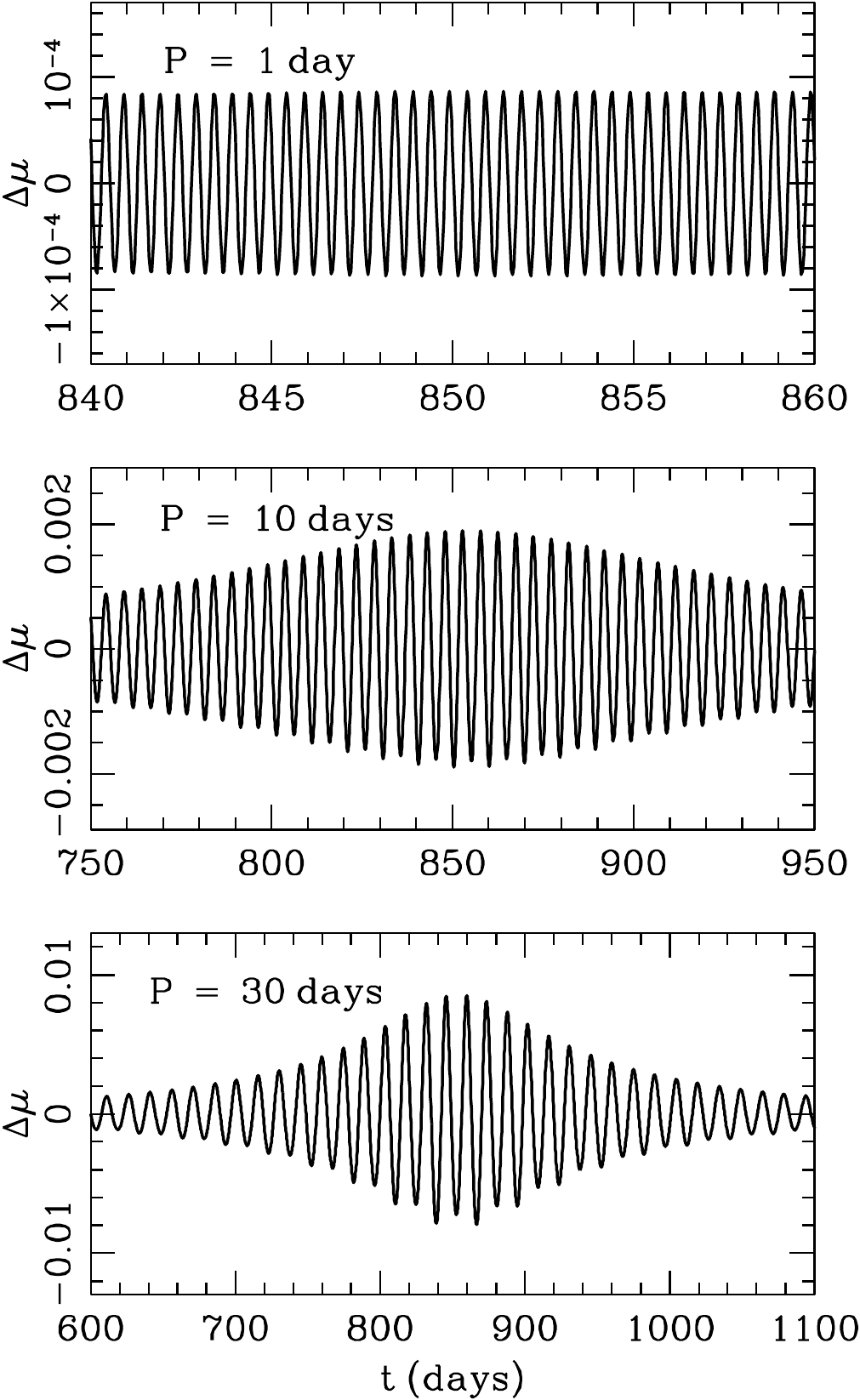}
\vspace{.2 true in}
\caption{\label{lcresiduals} The three panels show the residuals from
subtracting point mass lightcurves from the lightcurves produced by
the binary lenses with periods shown as red stars in Figure
\ref{prange.small}.  Other system parameters are the same as for
Figure \ref{lcperiods}.} 
\end{figure}

\begin{figure}[h]
\epsscale{0.5}
\plotone{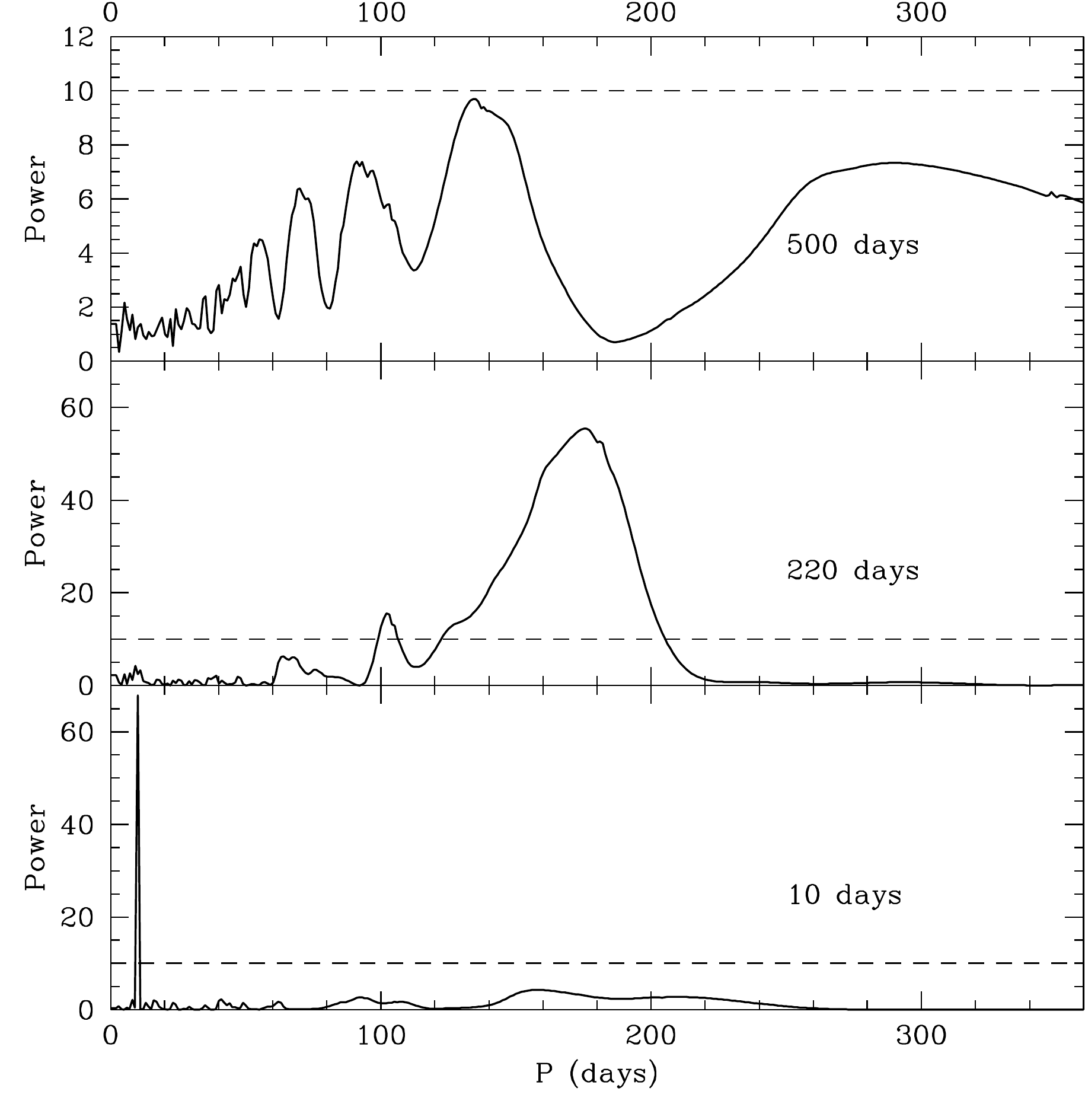}
\vspace{.75 true in}
\caption{\label{lcpower} The results Lomb-Scargle periodogram analysis for 
three of the lightcurves shown in Figures \ref{lcperiods} and \ref{lcresiduals}. 
The actual periods of the binaries are indicated in the figure.  Each 
lightcurve was sampled randomly roughly every 2 days and an uncertainty of 
$1\%$ and $10^{-3}\%$ was added to the data for the two longer period 
binaries and the short period binary, respectively.  The dashed lines show 
the power corresponding to 98\% confidence limit for the detection of 
periodic variability.}
\end{figure}

The middle panel in the top row 
shows the binary-lens light curve for $\alpha= 0.3$. 
The larger value of $\alpha$ yields larger deviations from 
the non-rotating light curve.  The timescale ratio $\tau_E/P$ is smaller 
for this system, so the translational motion of the binary center of mass 
has a stronger distorting effect on the rotational signature.  Indeed, an 
orbital period of $123$~days recovered by the 
Lomb-Scargle analysis is $12\%$ lower than the true period of $140$~days.  
This trend continues with the third panel in the first row, for which
$\alpha=0.4,$ producing an orbital period of $220$~days, just slightly
smaller than the value of $\tau_E.$  Here, too, the periodicity is
clear in the Lomb-Scargle periodogram, but the derived value is
$175$~days, $20\%$ lower than the actual value.  

The light curves in the bottom row of Figure \ref{lcperiods} are generated by 
binaries with
larger values of $\alpha,$ and, consequently, longer orbital periods.
In each case, $P_{orb} > \tau_E$ and the periodogram analysis
does not show a significant periodicity. Nevertheless, the
rotational effects are large enough that the light curves for rotating 
binaries are significantly more complex than their
corresponding static-binary counterparts, except possibly for the example with
$\tau_E/P=0.07$.  

The red dots in Figure \ref{prange.small} all correspond to values of 
$\alpha$ smaller than
$0.2.$ In these cases, the deviations from  the point-lens form are
very small and would likely not be detected in the absence of orbital motion. 
The periods are short enough ($30$~days, $10$~days, $1$~day), 
however, that there are many orbital revolutions during a typical lensing 
event. As a result, the periodic nature of the photometric fluctuations is 
very obvious, as long as they can be detected.  Figure \ref{lcresiduals} 
shows the residuals between the best-fit point-lens light curve in each case 
and the binary-lens light curve.
For the cases of $30$~days and $10$~days, millimagnitude 
photometry
would detect the deviations from the point-lens
form. For the 1-day orbit, photometric sensitivity
comparable to that of {\sl Kepler} or the upcoming {\sl TESS} mission
 would be needed. 
In all three cases,  the orbital period would be extracted by
the Lomb-Scargle analysis with a high degree of confidence and precision.  
 
Figure \ref{lcpower} shows the results 
for the Lomb-Scargle analysis (with the factor of 2 correction for the period,
as described above) of 
the three lightcurves taken from those 
shown in Figures \ref{lcperiods} and 
\ref{lcresiduals}.  For the longest period system
there is no convincing signal, since actual period of the binary here is 
more than half of the total duration of the event (which we have taken to 
be $3 \tau_E$, roughly corresponding to overall magnification above $2\%$).
By contrast, the middle panel shows the case where a 175-day period is 
detected with very high significance.  This result is expected, since 
for this system $3 \tau_E/P = 3.3$, so the binary undergoes over 3 
complete revolutions while the event is detectable.  However, the detected 
period is {\it not} equal to the binary period of the system.  This discrepancy
is due to the motion of the source with respect to the binary.  
In Guo et al. 
(2011) we discuss a modified timing analysis approach that allows us 
to extract correct periods for such systems.  
Finally, the lower panel shows 
that there is no problem extracting the correct period for the systems 
where the event duration is many times higher than the 
orbital period (since 
here the motion of the source is negligible over 
one orbital period of the 
lens).    


\section{Exploring a wide range of systems}

\subsection{Systems for which orbital phase changes 
influence lensing light curves}

In this section we display magnification maps and light curves
for a small number of binaries, but begin by showing 
that each represents a much larger set of binaries that can
generate light curves that are 
identical or else similar in all essential details
 to the ones
shown here.

The crucial issue for detecting evidence of orbital motion is the 
ratio of the time duration of the event, to the time required for a significant
change of phase to occur. We use the parameter $\tau_E/P$ as a 
proxy for this ratio, but the exact value of $\tau_E/P$ is not crucial,
because the duration of the detectable event may be longer or shorter
than $\tau_E$, while the required orbital phase change can be smaller than or
larger than $2\, \pi$. 
If, for example, the photometry 
and sampling are good enough to allow the detection of an event over long
intervals of time, or if the event is studied astrometrically,
then orbital effects can be significant for 
relatively values of $\tau_E/P$ of $10$ or more.

To study a variety of situations, we consider the $5$ systems (binaries
and planetary systems) listed in Table~1.
Together, these systems span
a range of mass ratios from unity to the planetary regime and of
primary masses from those characteristic of planets or brown dwarfs to those
associated with stellar-mass BHs. Although
in \S 2 we showed that both smaller and larger values for $\alpha$ can 
be associated with detectable rotation-induced deviations, here we
concentrate on a 
somewhat more limited range of $\alpha$.
\begin{table}[h]
\begin{tabular}{cccccccccc}
\hline
System & $M_1/M_{\odot}$ & $M_2/M_{\odot}$ & $a$\,(AU) & $P$\,(days)& 
$D_L$\,(pc) & $D_S$\,(pc) & $v_T$\,{\rm (km/s)} & $\alpha$ & $\tau_E/P$ \\
\hline
1 & 0.05 & 0.05   & 0.11 & 42  & 30  & 8000 & 15 & 0.7  & 0.86 \\
2 & 1.3  & 0.3    & 0.28 & 44  & 50  & 8000 & 10 & 0.35 & 6.4 \\
3 & 0.8  & 0.002  & 0.17 & 28  & 30  & 8000 & 15 & 0.38 & 3.6\\
4 & 7    & 2      & 1.67 & 263 & 200 & 8000 & 25 & 0.44 & 2.0 \\
5 & 0.006& 0.0006 & 0.016& 8.7 & 50  & 8000 & 15 & 0.3  & 1.4 \\
\hline
\end{tabular}
\caption{A summary of parameters for the five representative binary lens 
systems discussed in this section. We list primary and secondary masses, 
semi-major axis, orbital period, lens and source distances, 
relative transverse velocity $v_T$, $\alpha,$ 
and the timescale ratio $\tau_E/P$.}
\end{table}

\subsection{Extension to the wide range of systems represented by each individual binary} 

There are three ways in which a single system exhibiting 
the effects of phase changes can represent a larger range of 
different systems.

The first is that a new system can have the same
orbital parameters, be located in the same place relative to us, and have the 
same transverse velocity, but the physical density of the lens can be different.
Thus, a mass of roughly $1.3\, M_\odot$ may correspond to a main sequence star,
a giant star, a white dwarf, or a neutron star. Systems with
compact objects are common, because many lens systems
are old enough that  one or both components of a 
binary or higher-order multiple
have evolved, and roughly $25\%$ of these systems will have undergone
one or two epochs of mass transfer. 

The second way in which an individual system represents many others, is through 
the similarity of the light curves to those generated by systems 
with somewhat different parameters.
Considering, for example, the 
characteristics of the binary lens,  
 Figure~1 shows that 
the shape of the isomagnification contours and the positions of caustics
are determined by the binary mass ratio, $q$, and the separation between the
binary components, $\alpha$, expressed in units of the Einstein radius. 
As $q$ and $\alpha$ are slowly changed from the particular values given in
Table~1, the isomagnification contours also change slightly, generally morphing into 
shapes that will yield significantly different light curves only when
the values have been changed by more than about $10\%-20\%$. 

Similarly, the 
source path 
does not need to fine-tuned to produce light curves with similar
characteristics.
Light curves
similar in their essentials (e.g., number and spacing of peaks, 
maximum magnification, etc.) to each of the light curves we 
show could have been generated when a background source follows a
somewhat different path behind the lens. The distance of closest
approach could have been a bit smaller or larger and/or the angle of
closest approach could have been somewhat different. In general, there
is a set of paths that would have produced light
curves similar  
to the one we are studying. The details of the new set of
paths might be somewhat different in small ways,
but the size of the range should be comparable.
Neither is fine tuning
of the orbital phase required in order for the events
to exhibit clear evidence of effects of rotation. 
The orbital phase becomes less relevant when the orbital period 
is short relative to the duration of the lensing event.
As we will show in some of 
the examples in the next section, there is some difference between the
cases in which the motion is prograde, versus retrograde. By prograde, we
mean that the translational motion of the background source
is along the same direction
as the orbital motion.  
When the motion is retrograde, there are more opportunities for irregularities
in the magnification pattern to rotate in front of the  source path,
increasing the effects of rotation on the light curve shape. 

Finally, as discussed below, the symmetries in the basic equations mean that
it is possible to transform a particular system
into a very different one that will produce identical or nearly identical
light curves, although generally with a rescaling of the time.

\subsubsection{Similar Values of $\tau_E/P$ for Physically Different Binaries}

To demonstrate that a range of very different physical systems can
produce light curves almost identical to those that will be
shown for the binaries listed in Table~1, we conducted a 
computational experiment. 
Specifically, to find binaries with similar values of $\tau_E/P$,
we generated a large number of binary lenses.
Because the shape of the light curve depends on the value of $\alpha,$
we selected only binaries with 
 $\alpha$ having values 
within $0.05$ of the value shown in Table~1. 
Because the value of $q$ doesn't influence the orbital period,
we generated, in addition to $\alpha,$
 only the relative transverse speed $v_T,$ the
total mass, $M,$ the lens distance $D_L.$ 
We considered the direction toward the Galactic Bulge, and used
$D_S=8$~kpc.    

\begin{figure}[]
\epsscale{.75}
\plotone{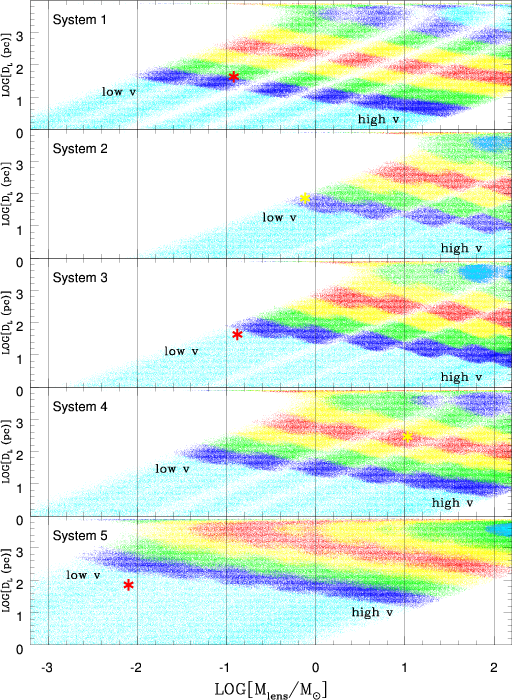}
\vspace{-.2 true in}
\caption{\label{all_sys} 
Points on each panel correspond to binaries that have the same value of
$\tau_E/P$, to within $0.2$, of the value listed for the system 
in Table~1 that labels the
panel. These binaries also have values of $\alpha$ within $0.05$ of the 
value listed for the system in Table~1. Points toward the left (low-velocity
side) have $v_T$ near $10$~km~s$^{-1}$; $v_T$ increases toward the
right, to a maximum of $\sim 100$~km~s$^{-1}$ on the right. Points in cyan near
the bottom have orbital periods, $P$, less than $25$~days. Proceeding upward,
points in dark blue have $25\, {\rm days} < P < 50\, {\rm days}$; 
points in green have $50\, {\rm days} < P < 100\, {\rm days}$. The lower
and upper limits of the interval of orbital periods increase by a factor of 
$2$ for each subsequent colored band.}
\end{figure}

Each point in Figure \ref{all_sys} represents a binary lens with $\alpha$ chosen as 
described above and with the value of $\tau_E/P$ lying within $0.2$ of
the value given for the corresponding system  in Table~1.  
With values of both $\alpha$
and $\tau_E/P$ very
close to the values used in Table~1, we are guaranteed that the 
binary we generated can produce a light curve almost identical to 
those shown 
in Figures~8 and 9 (System~1), 
Figure~10 (System~2 in the left-hand panel, System~4 in the
right-hand panel); Figure~11 (System~3); Figure~12 (System~5). 

The result is that a light curve with the same morphology can be produced by lenses 
having a wide range (several orders of magnitude) of total masses, 
of distances to the lens, orbital periods, and transverse speed. Furthermore,
many portions of the regions covered by  
colored points correspond to real physical systems at
various stages of evolution.  
To derive the points in Figure \ref{all_sys}  this, 
we varied the parameters as described below.

\smallskip

\noindent{\bf Relative transverse speed, $v_T$:} The value of 
$\tau_E/P$ is inversely proportional to $v_T$. The value of $v_T$ 
therefore plays an important role in determining whether
the effects of the lens system's orbital motion can be detected.
We generated $7$ small ranges of values of $v_T:$ $9-11$~km~s$^{-1}$,
$14-16$~km~s$^{-1}$,   
$24-26$~km~s$^{-1}$,   
$39-41$~km~s$^{-1}$,   
$59-61$~km~s$^{-1}$,   
$79-81$~km~s$^{-1}$,   
$99-101$~km~s$^{-1}$.
In each panel of Figure \ref{all_sys}, points on the left-hand side (labeled ``low-v'')
have values of $v_T$ near $10$~km~s$^{-1}$,   
and those on the right-hand side (labeled ``high-v'')
have values of $v_T$ near $100$~km~s$^{-1}$. 
The distribution of points in each panel shows 
evidence of the transitions between the seven velocity ranges. 
      
\smallskip 

\noindent{\bf Lens mass:}  
We selected the logarithm (to the base 10) of the mass to 
uniformly populate the interval between
$-3.2$ (corresponding to 
approximately $0.6$ the mass of Jupiter) and $2.2$ (corresponding to 
approximately $158\, M_\odot$, close to the mass of the most massive star
known).  
Bright stars, whether they are high-mass stars on the main sequence
or giants, produce fewer detectable events because they are will outshine
most background stars they lens by several magnitudes. Thus, the
more massive stars likely to produce detectable events are 
stellar-mass BHs. The masses of the known stellar-mass BHs 
extend to roughly $35\, M_\odot$ (Prestwich et al. 2007) and the
evolution of massive stars (as well as possible BH mergers) may produce even more massive BHs.

\smallskip 

\noindent{\bf Distance to the lens:}
The functional form of $\tau_E/P$ exhibits a symmetry between $D_L/D_S$
and $(1-D_L/D_S)$. Thus, for a fixed $D_S,$ the value
of $\tau_E/P$ is the same for a lens $100$~pc from us
as it is for a lens $100$~pc from the source star.  If, therefore,
we had 
plotted $D_L$, instead of $log_{10}(D_L)$, 
each panel of Figure~7 would reflect a symmetry about the axis defined by
$D_L=D_S/2.$ To generate the points in these panels, we 
 selected the logarithm (to the base 10) of $D_L$ to lie between $0$ 
and $3.99$\footnote{Note that, to better fill in the region of $D_L \approx D_S$
we could have continued sampling to $log_{10}(D_L) = 3.99996,$ but this is
not necessary here, since the upper region is not well resolved
when the logarithm of $D_L$ is plotted.} 

Note that, although there is a formal symmetry, there are some differences
between events generated by nearby lenses and those generated by distant
lenses. Nearby lenses are more amenable to direct detection. This can permit
a wide range of post-event observations that provide more information about the 
binary. In addition, parallax effects are more likely to be
detectable in the light curve, and should be included in the model fits. Parallax
 provides information about
the distance of the lens from us. On the other hand,
distant lenses will be detectable only if they are
bright and/or happen to be in less crowded fields. In addition,
when the lens is more distant, 
finite-source-size effects are more likely to affect the light curve shape, especially
when the lens is a binary whose light curve displays short-lived features.
Such effects can play two opposing roles in our study. 
If the source size is large  compared to 
the binary features exhibited by the isomagnification contours,
finite-source-size effects can diminish the effects of binarity in the light curve
and wash out some of the short-lived deviations. On the other hand, if the
source size is smaller, but still detectable, finite-source-size
effects can be used to measure the size of the Einstein ring, providing
a relation between the lens mass and its distance from us.

The results are shown in Figure \ref{all_sys}, in which 
the label of each panel indicates that the panel 
corresponds to one of the systems listed in Table~1.
Each point in a specific panel represents a binary lens with $\tau_E/P$
within $0.2$ of the value shown for this system in Table 1. 

\subsection{The light curves of the sample systems} 

\subsubsection{System 1}

\noindent{\bf The specific lens system considered: two low-mass companions.} 
The system parameters we have selected correspond to
a brown-dwarf/brown-dwarf binary, with equal mass components.
The Einstein angle is approximately equal to $5$~mas, and the
Einstein radius is $0.155$~AU. Thus, the orbital separation is
roughly $0.7\, R_E.$ Of special interest to our investigation is that
the Einstein diameter crossing time is somewhat smaller than the orbital
period.  

Such binaries are likely to be quite numerous.  In fact, 
one of the closest stellar systems to 
the Sun (at $3.6$~pc) consists of two  
T dwarfs, with $M_1\sim 0.05\, M_\odot$ and $M_2 \sim 0.03\, M_\odot$.
These components are separated from each other by $2.1$~AU. 
This low-mass binary it itself in orbit around the dwarf star Epsilon Indi,
with an orbital separation of  
orbiting at a distance 
roughly $1500$~AU
(McCaughrean et al. 2004; Scholtz et al. 2003, Volk et al. 2003).  
The presence of a double brown dwarf so
close to the Sun indicates that such systems are not likely to
be rare.   Indeed, based on recent brown dwarf search programs, the space 
density of T dwarfs in the solar neighborhood is estimated to be of order 
$\sim 0.005-0.05\,{\rm pc}^{-3}$ (see e.g. Geissler et al. 2011, 
Kirkpatrick et al. 2011), which implies the presence of 2500 to 25000 of these
objects within 50\,pc from the Sun.  A significant fraction 
is expected to be in binaries with 
companions less massive than $0.1 M_{\odot}$ (Geissler et al. 2011), i.e. 
systems very similar to that we consider here.

Thus, the system we consider here may be viewed as a proxy for the
general class of nearby low-mass binaries. 
Physical systems located within a few tens of pc and
capable of producing similar light curves to those computed
for System~1 include a wide range of low-mass dwarf binaries,
some of which could have mass ratios as low as $\sim 0.8$: 
(1)~low-mass stellar binaries. (2)~binaries where one component
is a low-mass star and the secondary is a high-mass brown dwarf.
(3)~planetary systems where one component in a brown dwarf and
its companion is a high-mass planet.

\noindent{\bf The light curve:}  
System~1 has $\tau_E/P = 0.86,$ the smallest value of considered in 
this section. The system will have executed $2-3$ orbits
during the time the magnification is detectable, assuming that 
deviations from baseline of $10\%-1\%$ can be detected.  
System~1
was also chosen to have values of $q$ and $\alpha$ that are most
likely to produce detectable binary effects in the light curve, even in the
absence of rotation. $\alpha=0.7$ stretches the isomagnification
contours, enhancing binary detection. This effect would be even more
pronounced for larger values of $\alpha$.  
The magnification pattern is symmetric, and this means that there
are two opportunities to pass through regions with higher magnific   
Light curves generated by this lens are displayed in
Figure~9 and Figure~10.  
\begin{figure}[h]
\epsscale{1}
\plottwo{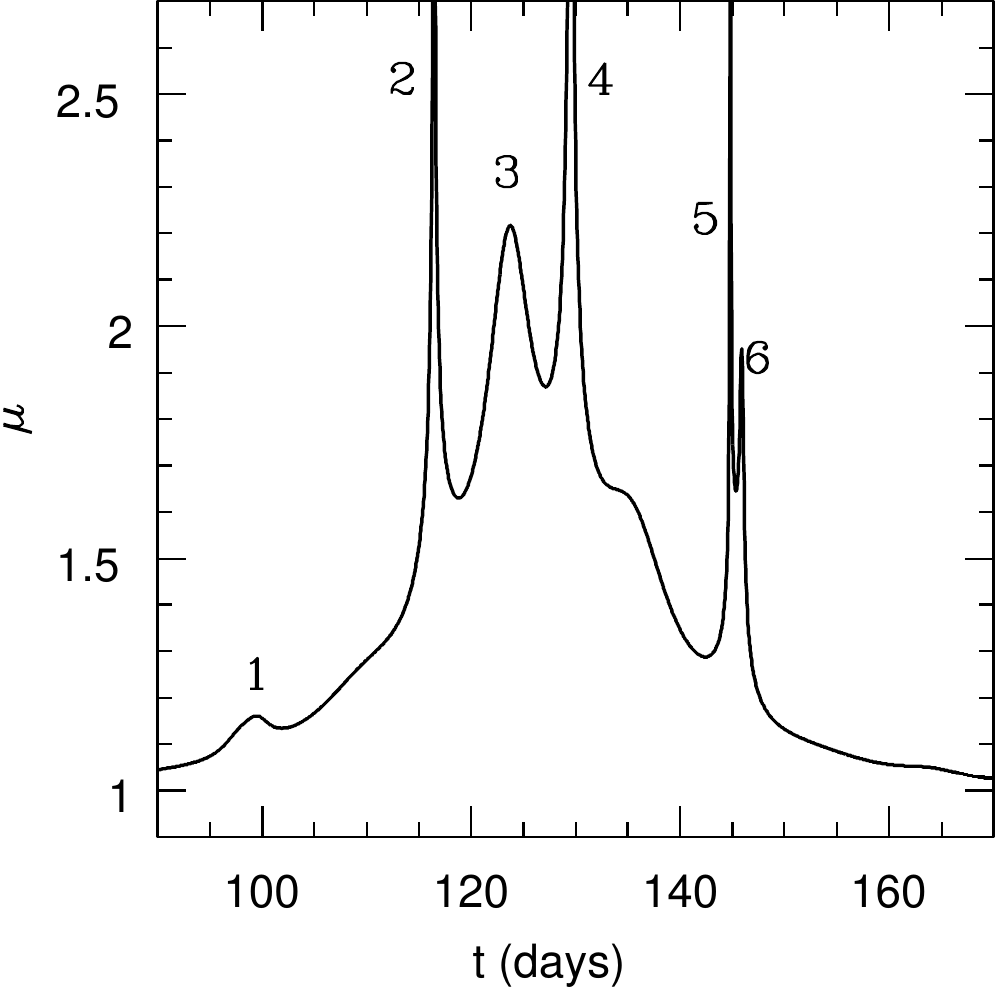}{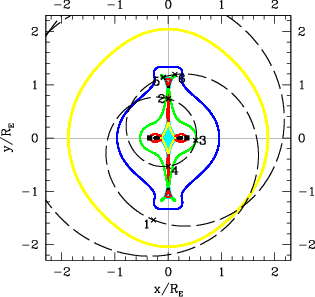}
\vspace{-.15 true in}
\caption{\label{sys1} {\bf System 1.} Panel (a) shows a lightcurve 
calculated for $b=0.5$.  The magnification map for this binary is
displayed in panel (b).  
The two black
squares on the $y=0$ axis mark the location of the two stars.  The
path of the source star, moving clockwise in the binary reference
frame, is drawn as a dashed black line.  The numbered black crosses
mark the locations along the trajectory which produce the strongest
features in the lightcurve in panel (a).  Note that rotation of the
lens system produces three caustic crossings for a relatively large
impact parameter event.}  
\end{figure}
\begin{figure}[h]
\epsscale{1}
\plotone{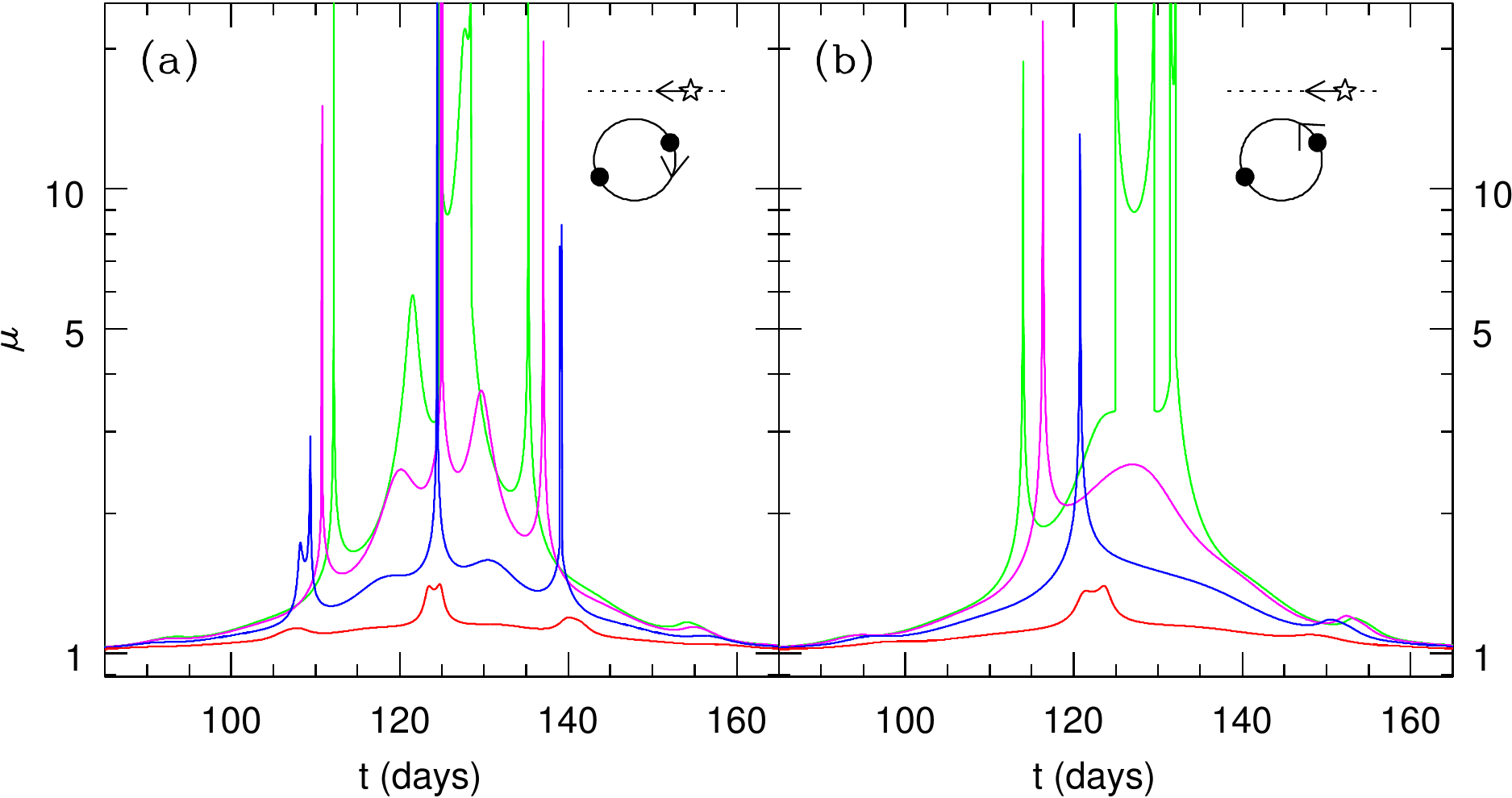}
\vspace{0 true in}
\caption{\label{sys1.lc} {\bf System 1:} A sequence of microlensing lightcurves for System 1
calculated for $b=0.1,\,0.3,\,0.7,\,1.3$, in order of decreasing
magnification.  The two panels illustrate the difference between
retrograde (panel a) and prograde (panel b) geometry, as illustrated
Panel on the left shows the results for a binary rotating clockwise
(retrograde), while the one on the right rotates counterclockwise
(prograde).}
\end{figure} 

Figure \ref{sys1} illustrates the effect of binary rotation on a 
relatively high impact parameter microlensing lightcurve for System~1.  
Panel (b) shows
the path of the source in the fixed frame of the binary.  
It appears to circle 
the binary center of mass clockwise, completing roughly two
revolutions while the overall magnification remains above 1.05.  
The spiraling
source path insures that a much larger portion of the lensing plane is 
seen by the source. As a result there are many more strong features in the 
resulting lightcurve than can be expected for a typical binary lightcurve 
for which rotation does not play a large role.  
The repetitive nature of these 
features suggests the effects of orbital motion,
and the standard Lomb-Scargle analysis of the lightcurve does identify the
signal as periodic (though the false signal probability is just over 2\%, close
to our threshold).  The inferred period is 45\,days, close to the actual value
of 42 days. 

Because it increases the fraction of the magnification plane explored 
in each lensing event, orbital rotation significantly increases 
our chances of 
detecting binarity for large impact parameter events.  
Figure \ref{sys1.lc}\, (a) 
shows a sequence of lightcurves for four different impact parameters.  Even 
for a distant approach, $b=1.3$, 
there is a hint of a periodicity in the three magnification
bumps apparent in the lightcurve.   This sequence also underscores the 
need for relative motion correction in determining the period 
via the timing analysis.  The periodicity 
is very apparent in all of the curves, but it appears 
that the underlying period appears different for each one.  
 
A comparison between panels (a) and (b) in 
Figure \ref{sys1.lc} emphasize  the 
importance of the direction of binary rotation with respect 
to the lens-source 
relative motion.  The cartoons in the upper-right-hand corner in each panel 
illustrate our definitions of prograde and 
retrograde rotation; note that the binary center-of-mass 
is assumed to be fixed, while the source moves past the binary.  
The retrograde motion (panel a) corresponds 
to the case when the companion closest to the projected position 
of the source moves opposite 
to the direction of source velocity.  In this 
situation, the effects of binary rotation are amplified 
and the lightcurves 
show more peaks on average.  The opposite is true for prograde rotation 
(anel b), when the closest companion and the 
source move in the same direction, 
partially canceling the effects of rotation.  As a result, the prograde 
lightcurves in panel (b) look more like the ``standard'' binary 
microlensing events than those in panel (a).

\subsubsection{System 2}  

\noindent {\bf The specific lens system considered: A
nearly solar-mass binary.}
System 2 can represent a common type of binary 
in which both stars are on the main sequence. The mass ratio we chose is 
similar to the value at the peak of the observed distribution 
of binary mass ratios measured by Duquennoy \& Mayor (1982). The 
orbital separation 
is not unusual, although the observed distribution peaks at somewhat larger 
values.  

Alternatively, System 2 
represents a binary in which one or both components have
already evolved. In fact the smaller mass ($0.3\, M_\odot$) and the 
orbital period ($44$~days) satisfy the period/core-mass relationship
(Rappaport et al. 2005).
This relationship applies to a situation in which a giant star filled its
Roche lobe, transferring mass to a companion until the time at which
the giant's stellar envelope was depleted. In this case, the depletion of the
envelope left behind a helium WD with a mass of $0.3\, M_\odot.$
There are two possibilities. This core could be the remnant of the star
that was initially the most massive. In this case the $1.3\, M_\odot$
star is still on the main sequence. Since it would have gained mass from the
giant's envelope, it can be viewed as a ``blue straggler'', with a 
larger mass than its initial mass. On the other hand, the core could
be the remnant of the initially less-massive star. In this case the
$1.3\, M_\odot$ star has already evolved
and is either a WD or a NS. The compact object is also likely
to have gained mass from the giant that evolved to become the $0.3\, M_\odot$ 
heium WD. If today's more massive star is a WD, then its mass
is close to the Chandrasekhar mass and it is a near-miss Type Ia supernova. 
If instead it is a neutron 
star, it may have been spun up  
by mass accretion and may therefore be a millisecond pulsar.
Thus, this single example, one of many that could have been chosen, illustrates
that important classes of evolved stellar binaries can imprint signatures
 of rotation onto lensing light curves.

\noindent{\bf The light curve:} 
The left-hand  pane; of Figure~10 shows two light curves generated by this lens;
as before, the source is assumed to 
be located near the Galactic center.  The timescale
ratio is $\tau_E/P = 6.4$,  so the binary goes through many  
full rotations during the duration of the event.   
The resulting light curves 
have a highly periodic structure with the source returning over and over 
to roughly the same regions of the lens magnification plane.  
It is typical that 
the increasing numbers of features are associated with
short orbital periods and small values of $\alpha$, so that the 
amplitude of the deviations is relatively small  
(see Figure \ref{lcperiods}).  It is easy to understand this 
by noting that  
the timescale ratio depends inversely on $\alpha$ (see Eq. [\ref{ratio}]).  
Indeed, for this system $\alpha = 0.35$, which is half of its
value for the brown dwarf binary we discussed in the previous section.  
Smaller
$\alpha$ implies that, over long time scales, the binary lensing features are less prominent, so 
overall the lightcurves look more like those due to single-mass lenses.  

\begin{figure}[h]
\epsscale{1}
\plottwo{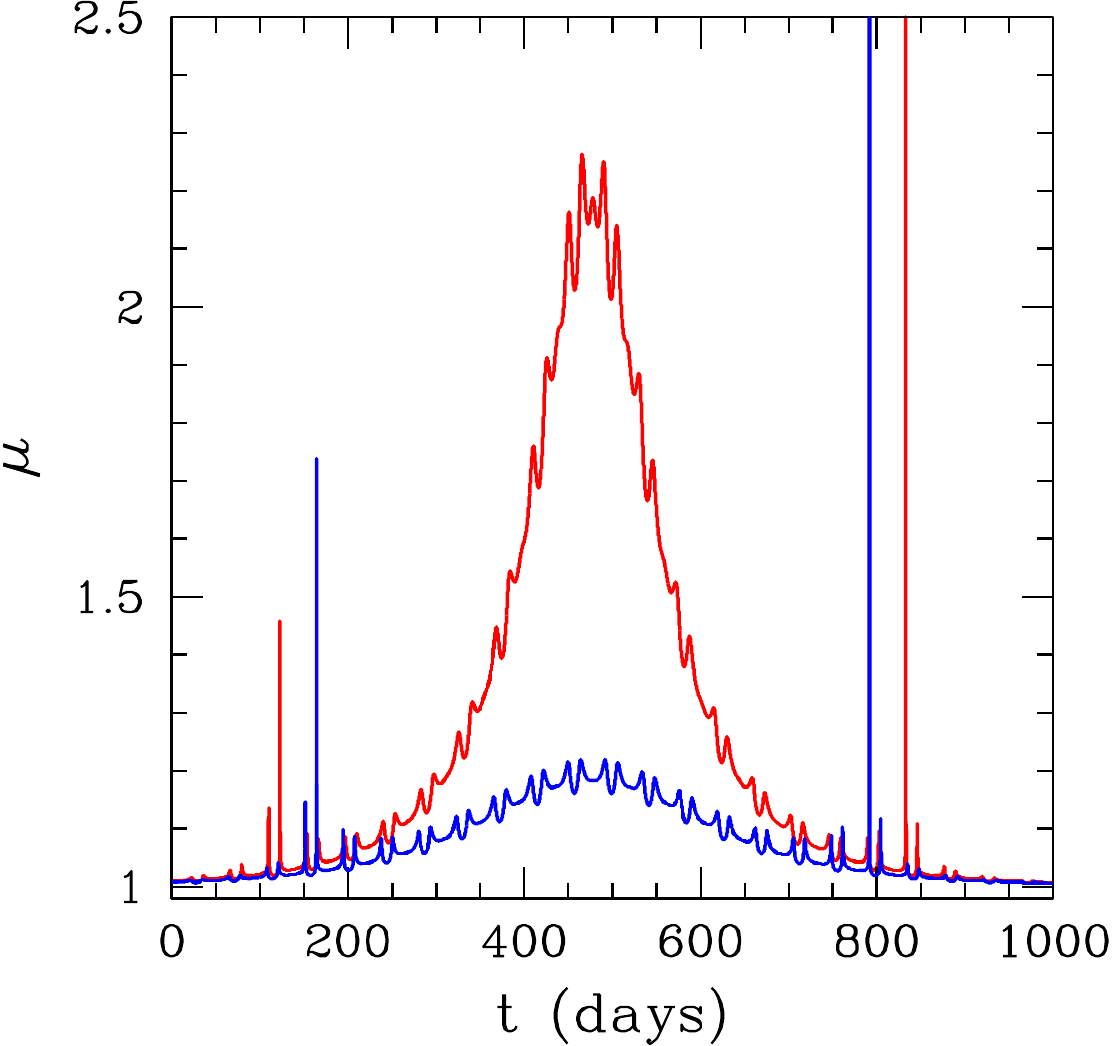}{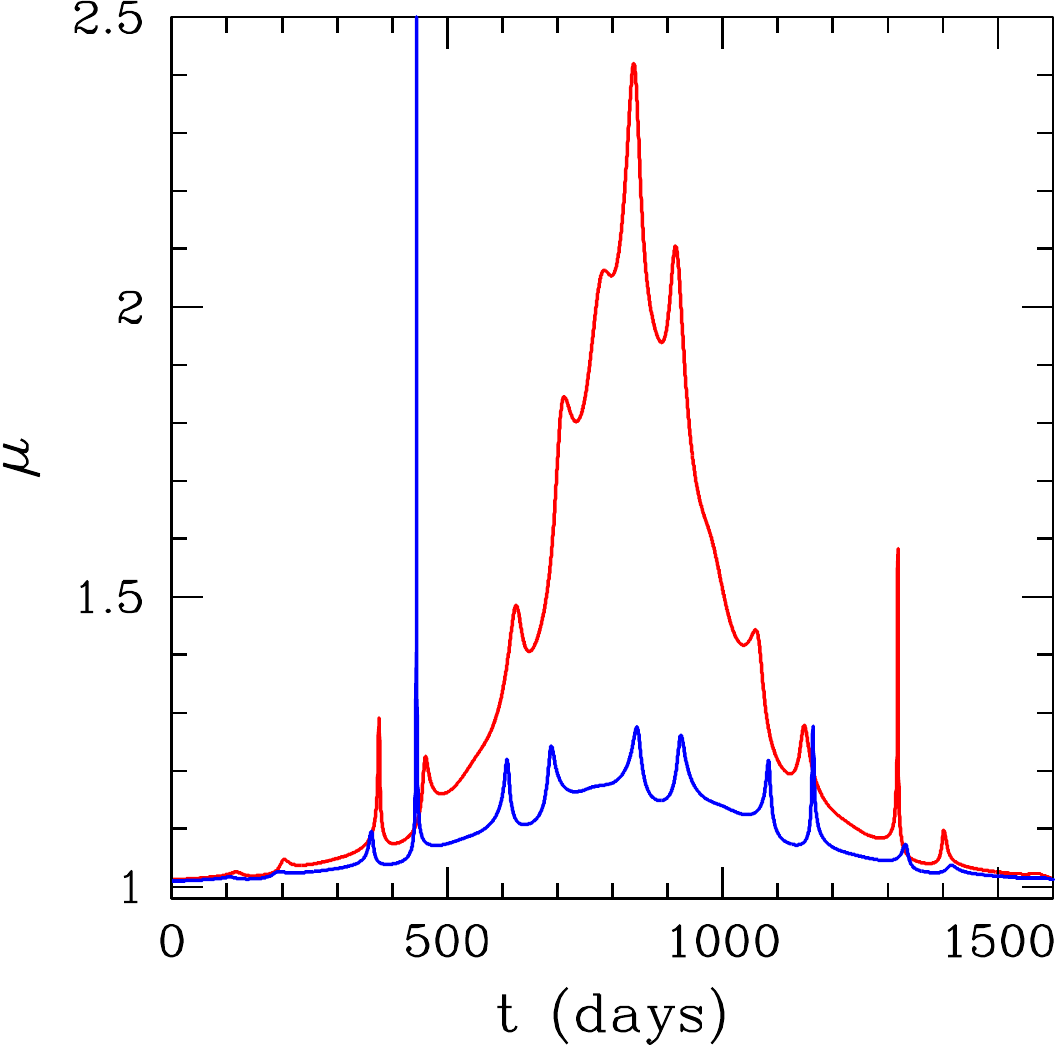}
\vspace{0 true in}
\caption{\label{sys2.sys4} {\bf System 2 and System 4:} Panel (a) shows a sequence of microlensing 
lightcurves for System 2 calculated for $b=0.5,\,1.3,\,2.0$, in order of 
decreasing magnification. Panel (b) shows lightcurves for System 4
corresponding to $b=0.5,\,1.3$.}
\end{figure}

Unsurprisingly, the Lomb-Scargle analysis for both of the lightcurves shown 
in Figure \ref{sys2.sys4}(a) extracts the orbital period with a 
discrepancy from the true orbital period smaller than $4\%$, 
with a very high degree of confidence. Since 
the timescale ratio is large, the corrections for the relative 
lens-source motion are not very important in this case.

\begin{figure}[h]
\epsscale{1}
\plottwo{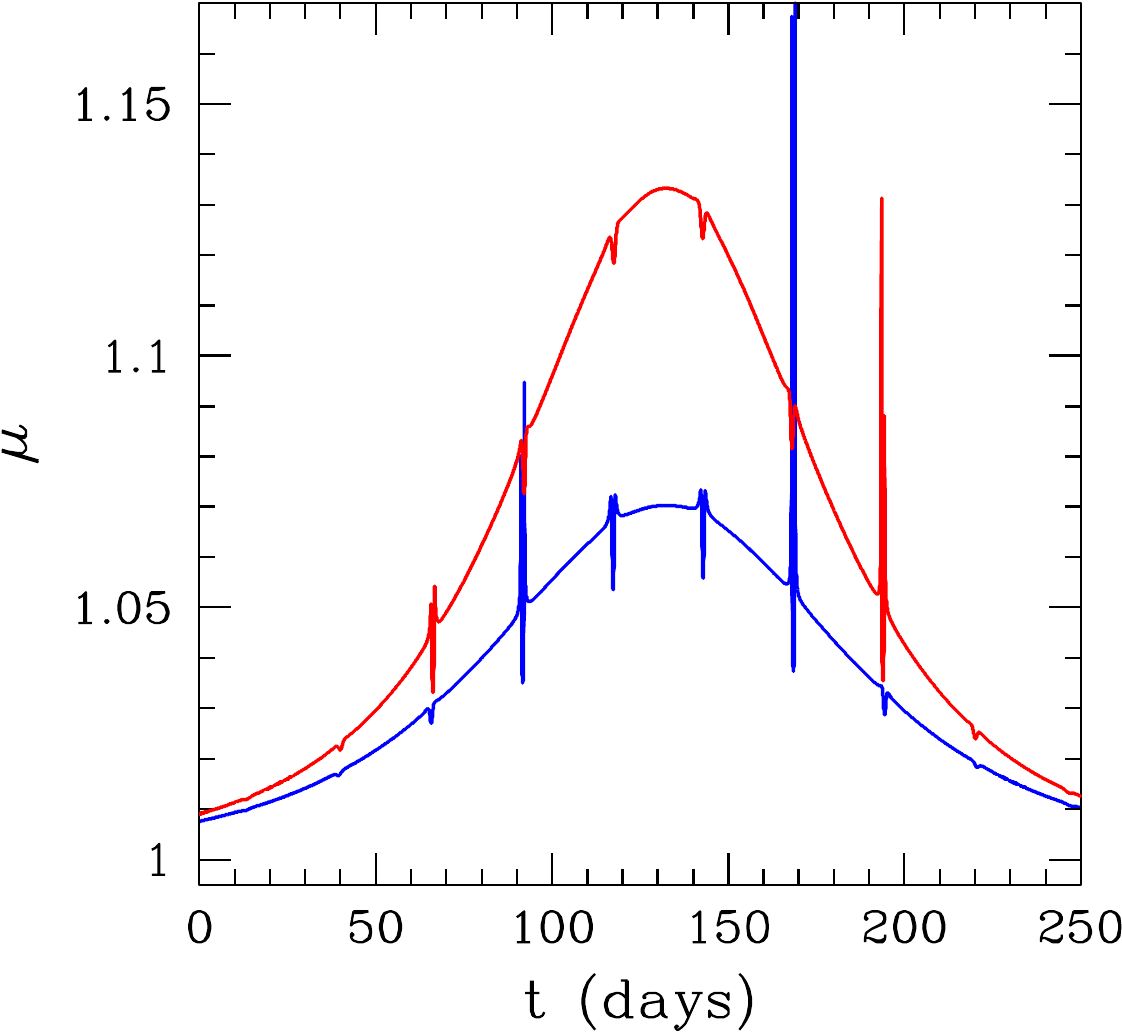}{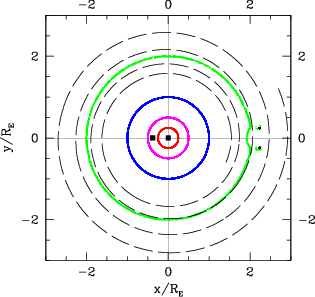}
\vspace{0.0 true in}
\caption{\label{sys3} {\bf System 3:} Panel (a) shows lightcurves for System 3,
calculated for $b=1.5,\,2.0$.  The magnification map for this binary in
displayed in panel (b).  
 The black curve is the path of the 
source in the binary frame which corresponds to the red lightcurve in panel 
(a).  Only the inspiral part of the source path is shown for clarity.}
\end{figure}

\subsubsection{System 3} 

\noindent {\bf The specific lens system considered: Jupiter Orbiting a Dwarf Star.}
Planets with masses comparable to or larger than that of Jupiter are common,
and many orbit solar-mass stars with periods of days.  
(Di\thinspace Stefano 2012) showed that orbital motion can be detectable. 
The maximum change in magnification from the point-lens form occurs when
the distance between source and lens is equal to $1/\alpha-\alpha$. The 
position of the perturbation is a function of the orbital phase, 
providing a periodic signal when the Einstein crossing time is long enough. 
We consider a stellar mass of $0.8\, M_\odot;$ the
planet's mass is $2\, M_J,$ yielding $q=0.0025.$

\noindent{\bf The light curve:} 
Figure \ref{sys3}\, (a) shows two light curves for System~3.
The distances of closest approach are  $b=1.5$ and
for $b=2$ for the upper and lower light curve, respectively. 
The light-curve deviations are regular but small in amplitude.
A Lomb-Scargle test fails to pick up the orbital period in this case.  This is
mainly due to the low amplitude and small duration of the observed features (in
the context of our adopted once-per-day lightcurve sampling with 1\% 
photometry errors), 
as well as the oscillatory behavior of the magnification within the features. 
However, an observation with sufficiently high cadence and sensitivity would
allow the period to be measured directly, since the timescale ratio 
($\tau_E/P = 3.6$) is large.  Merely folding the lightcurves to match 
the sharp peaks yields the period of 26\,days, very close to the true orbital 
period of 28\,days for this system.  
Since the deviations in the isomagnification diagram 
are very asymmetric (due to the very low mass ratio for this system), we 
expect roughly one deviation per orbit, as opposed to two expected when $q$ is 
of order unity.  So the factor of 2 correction we routinely apply in our 
period analysis is not appropriate here.

An interesting point to note is that
this type of planet-induced deviation can be detected even when the
distance of closest approach between the source and lens is large, and the 
peak source magnification is low.
Figure \ref{sys3}(b)  shows the source path relative to the lens system.
The deviations from the point-lens form do indeed occur when the distance
between source and lens is $1/\alpha-\alpha \sim 2.25$.
The case of close-orbit planets is discussed in more detail in
Di\thinspace Stefano 2012.     


\subsubsection{System 4}

\noindent {\bf The specific lens system considered: A High-Mass Binary.}
When the lens is a high-mass star, $\tau_E$ can be large
because the associated value of
$R_E$ is larger for any given $D_L.$
Furthermore, for a given value of the semimajor axis, the period is
shorter. The overall effect can be to produce large values of 
$\tau_E/P_{orb}$. 
If the primary of such a lens is a main-sequence star, 
the light curve is likely to
be affected by blending, making the detection more difficult unless the 
lensed source is also very bright. Nevertheless, some such events should be
detectable. 

Perhaps the case of greatest interest is that in which the
most massive component is a black hole. 
If the primary star in System 4 is a BH, and if the secondary is on the 
main sequence, the evolution of the secondary will eventually cause it to
donate mass to the BH, probably through winds in this case, 
making the system detectable as an X-ray binary. 
System 4 is 200\,pc 
from the Sun, 
but exactly the same results would be obtained for an identical 
binary located at $D_L = 7.8\,{\rm kpc}$.

\noindent{\bf The light curve:} Two lightcurves for this system are
shown in Figure \ref{sys2.sys4}(b).  The periodicity is clearly 
evident in the data and is picked up by the Lomb-Scargle analysis with very 
high significance.
However, the timescale ratio is low enough that the relative motion corrections
become important.  The inferred periods are 161\,days and 230\,days for 
$b=1.3$ and 0.5, respectively, significantly different from the true 
orbital period of the system. The long-lasting low magnification light curve 
could easily be mistaken for a variable star.

\subsubsection{System 5} 

\noindent{\bf The specific lens system considered: binary planets.}
The class of free-floating planets may include binaries; these  
would be difficult or impossible to discover {\it except through their 
action as gravitational lenses}. For discovered systems close enough to us,
direct imaging could be employed to verify the nature of the system.  
In addition, 
planets orbiting stars could have moons with masses in the range
of planetary masses.  
Figure \ref{prange} shows lens masses that decrease from the bottom panel
to the top panel.
Extrapolating to even lower masses, i.e., to the 
planetary regime, 
we expect values of  
$\alpha$ close to $\sim \alpha = 0.2$, 
implying small amplitude deviations from the point-mass form.
For our System~5 we choose a fairly massive binary in which one 
companion is 6 times more massive than Jupiter while the mass of the other is 
2 times larger that that of Saturn.

\begin{figure}[h]
\epsscale{1}
\plottwo{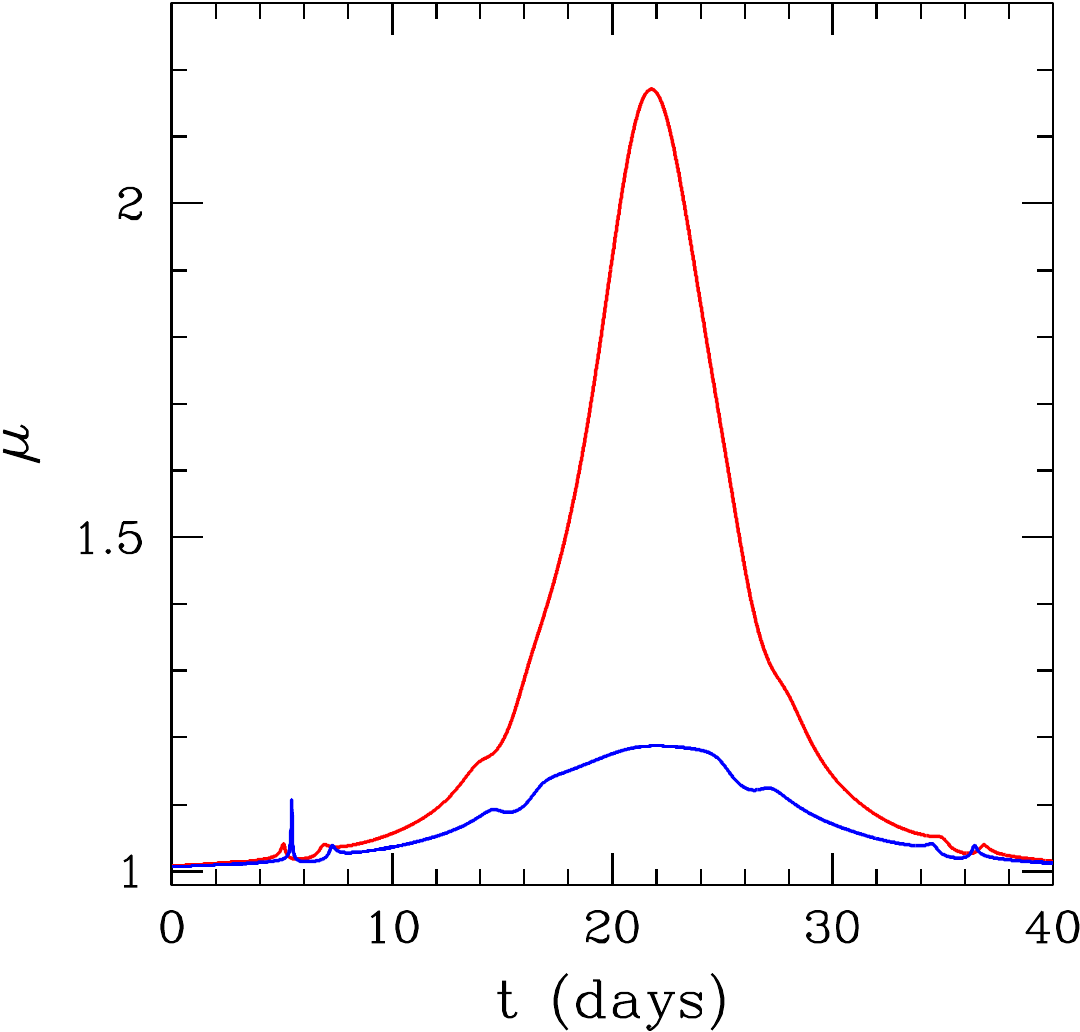}{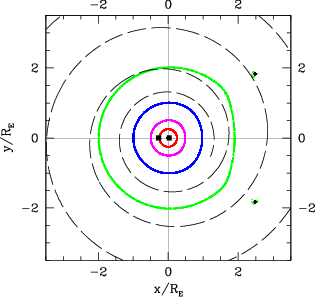}
\vspace{0.0 true in}
\caption{\label{sys5} {\bf System 5:} Panel (a) shows light curves for System 5,
calculated for $b=0.5,\,1.3$.  The magnification map for this binary in
displayed in panel (b).  The black curve shows the path of the source in 
the binary frame corresponding to the blue lightcurve in panel (a).  Only 
the inspiral part of the path is shown for clarity.}
\end{figure}

\noindent{\bf The light curve:} 
Figure \ref{sys5}\, (a) shows two representative 
lightcurves for this binary, while the isomagnification diagram with the 
source path for the lightcurve with $b=1.3$ are displayed in panel (b).  
Note the similarity between the isomagnification contours for this 
system and System 3,  though the mass ratio here is not as extreme, producing 
less asymmetry in the deviations from the point-lens form.  As expected, the 
observed deviations are relatively small (a few percent), but unquestionably 
periodic with an apparent period of $\sim 10\,{\rm days}$.  As was the case 
for System 3, Lomb-Scargle analysis does not 
pick up a believable period here, but one can easily be seen with improved 
photometric sensitivity.  Again, as in System 3, low value of the mass ratio 
makes the interpretation of the result difficult, since correcting by a 
factor of 2 only confuses the matter.  In addition, $\tau_E/P = 1.4$ means 
that relative motion corrections are important. 

\section{Conclusion}

We have shown that orbital motion produces significant phase shifts 
during gravitational lensing events 
for multiple lenses with wide ranges of parameters and distances from
us. This seems to be
 an odd result, because phase changes during lensing events have 
needed to be invoked only a small number of times, and the changes have
been relatively minor. 
The reason may be that many of the expected deviations are difficult
to identify. Some light curves exhibiting rotational effects
may be extreme enough that the event is not recognized as being due
to lensing or else is difficult to fit. More often, the effects associated with
orbital motion are subtle, 
and their discovery 
requires temporally dense, high-sensitivity sampling. Since different fields
are monitored in different ways, only about $10\%$ of
the events being discovered at present receive monitoring that is
adequate for these purposes. 
The detection efficiency may be low for many types of events 
(Penny et al. 2011b).

There are, however, some events for which the orbital-motion-detection 
efficiency should be
close to unity: events caused by short period lenses. For example,
if a Neptune-mass or Jupiter-mass planet orbits a solar-mass star
at $\sim 0.3\, R_E,$ 
then even distances of closest approach as large as
 $2.5-3\, R_E$ could produce distinctive, almost-periodic
deviations (Di\thinspace Stefano 2012). These would occur during a
stellar lens event that produces a peak magnification
 of only $\sim 1-3\%$.
The signatures of the planet
 could be easily missed without a filter that tests 
for intervals of periodicity in otherwise stable light curves. 
With such a filter, however,
detection of the signature is guaranteed for 
a wide range of mass ratios and with sampling as good as the present-day
sampling already applied 
in several fields today (Di\thinspace Stefano 2012). 
Thus, 
if $1/2$ of all stars have planets (a very conservative estimate), and if $1/2$
of these have nested planets, and if $1/2$ of these have planets with
orbital separations between $0.25\, R_E$ and $0.5\, R_E$, and if the mass
ratios are larger than $\sim 1/10^5$ in $1/4$ of these cases,  
then roughly $3\%$ of lensing events by stars would exhibit small signatures
of the planet's presence for approaches between $2\, R_E$ and $4\, R_E$;
in many cases (depending on the relative velocities), the signatures will 
repeat. 

A similar argument applies to binary systems, where orbital
distributions are typically modeled as being uniform in logarithm over $8$
intervals. If $1/2$ of all stellar systems are binaries, and the
$1/16$ of the stellar
primaries have a companion in an orbit  between $0.25\, R_E$ and $0.5\, R_E$,
this also suggests that $3\%$ of all stellar systems would have a 
significant probability of exhibiting orbital motion during lensing events.
Higher order multiples, which constitute $\sim 10\%$ of all stellar multiples
are even more likely to have one companion in an orbit for which 
orbital motion can be detected.

If, therefore, signatures of orbital motion 
are detectable in only $10\%$ of events discovered
today, we might expect $.006$ of  
the $2000$ events discovered per year to exhibit detectable evidence of orbital
motion. This corresponds to $12$ events per year, half of them 
planet-lens events. 
In addition, orbital motion could be discovered in
$42$ of the $\sim 14,000$ already-known events, 
$\sim 21$ of them associated with 
planet lenses.
Even if lower estimates of the population (Penny et al. 2011a) apply,
the numbers would still be significant, and will become more so with the advent
of programs such as {\sl KMTNet}. In addition,
the excellent astrometric
and
photometric 
precision possible with space missions 
({\sl GAIA}, {\sl Kepler,}
{\sl TESS} {\sl WFIRST}) 
will make detailed studies of orbital motion 
a regular part of microlensing discoveries.

Finally, orbital motion will play an important role in the study of
predicted gravitational lensing events. When the proper motion
of a star within a few hundred parsecs has been measured, we can check whether
it has a high probability of passing close enough to a background star to
cause a lensing event (see, e.g., L\'epine \& Di\thinspace Stefano 2012, 
Di\thinspace Stefano et al. 2013). The equations in Section~2 show that the
Einstein radii, $R_E$, of  nearby stars tend to be small. Thus, if the
orbital periods of planets orbiting these stars have a distribution 
similar to the distribution measured for the orbits of
already-discovered exoplanets, we expect that,
during lensing events, the orbital phase will 
shift significantly for close-orbit planets ($a< 0.5\, R_E$),
and often for planets in the ``resonant'' zone ($0.5\, R_E < a < 2\, R_E$),
and in some cases even for planets in wider orbits. 
Observing campaigns planned to monitor these predicted events will
 measure the masses of the central star and its planets,
determine the orbital period(s), and provide constraints on the orbital 
inclination. 
For many of these
systems, subsequent observations can verify the orbital properties and
can be used to learn even more. 
Planetary systems studied during observing campaigns
for predicted events will eventually
 become among the best studied exoplanetary systems.   
\bigskip

\noindent{\bf Acknowledgements:} It is a pleasure to acknowledge 
early contributions by Christopher Night and help from James Matthews
and Xinyi Guo. This work was supported in part 
by support from NSF AST-1211843, 
AST-0708924 and AST-0908878 and
NASA NNX12AE39G  AR-13243.01-A. 


\end{document}